\documentclass[lettersize,journal]{IEEEtran}
\IEEEoverridecommandlockouts

\usepackage{bbm}
\usepackage{amsmath}
\usepackage{bm}
\usepackage{mdframed}
\usepackage{amsthm}

\usepackage[utf8]{inputenc}
\usepackage{graphicx}
\usepackage[colorinlistoftodos]{todonotes}
\usepackage{amssymb}
\usepackage{longtable}
\usepackage[colorlinks=true, allcolors=blue]{hyperref}

\usepackage[noadjust]{cite}
\allowdisplaybreaks
\usepackage{lineno}
\usepackage{lipsum}
\usepackage{multicol}
\usepackage{subcaption}

\usepackage{multirow}
\usepackage{array}

\usepackage{xcolor}

\usepackage[vlined,linesnumbered,ruled,resetcount]{algorithm2e}
\SetArgSty{textnormal}

\usepackage{bbm}

\SetKwInput{KwInput}{Input}                
\SetKwInput{KwOutput}{Output}              
\SetKwInput{KwInitialize}{Initialize}                
\SetKwInput{myproc}{Procedure}{}{}

\usepackage{booktabs}
\usepackage{caption}
\usepackage{float}
\usepackage{capt-of}
\usepackage{arydshln}
\usepackage{soul,xcolor}
\usepackage{tabularx}

\usepackage{arydshln}
\usepackage[flushleft]{threeparttable}

\title{A Bayesian Framework of Deep Reinforcement Learning for Joint O-RAN/MEC Orchestration}

\author{\IEEEauthorblockN{Fahri Wisnu Murti, Samad Ali,  Matti Latva-aho}\\
\thanks{
Fahri Wisnu Murti, Samad Ali and Matti Latva-aho are with Centre for Wireless Communications, University of Oulu, Finland. 
This research has been supported by the Academy of Finland, 6G Flagship program under Grant 346208. 
}%
}

\begin{document}
\sloppy

\maketitle
\thispagestyle{plain}
\pagestyle{plain}
\begin{abstract}
Multi-access Edge Computing (MEC) can be implemented together with Open Radio Access Network (O-RAN) over commodity platforms to offer low-cost deployment and bring the services closer to end-users.
In this paper, a joint O-RAN/MEC orchestration using a Bayesian deep reinforcement learning (RL)-based framework is proposed that jointly controls the O-RAN functional splits, the allocated resources and hosting locations of the O-RAN/MEC services across geo-distributed platforms, and the routing for each O-RAN/MEC data flow. The goal is to minimize the long-term overall network operation cost and maximize the MEC performance criterion while adapting possibly time-varying O-RAN/MEC demands and resource availability.
This orchestration problem is formulated as Markov decision process (MDP). However, the system consists of multiple BSs that share the same resources and serve heterogeneous demands, where their parameters have non-trivial relations. Consequently, finding the exact model of the underlying system is impractical, and the formulated MDP renders in a large state space with multi-dimensional discrete action.
To address such modeling and dimensionality issues, a novel model-free RL agent is proposed for our solution framework. The agent is built from Double Deep Q-network (DDQN) that tackles the large state space and is then incorporated with action branching, an action decomposition method that effectively addresses the multi-dimensional discrete action with linear increase complexity. Further, an efficient exploration-exploitation strategy under a Bayesian framework using Thomson sampling is proposed to improve the learning performance and expedite its convergence. Trace-driven simulations are performed using an O-RAN-compliant model. The results show that our approach is data-efficient (i.e., converges significantly faster) and increases the returned reward by 32\% than its non-Bayesian version. Moreover, it outperforms Deep Deterministic Policy Gradient by up to 41\%.

\end{abstract}

\IEEEpeerreviewmaketitle

\begin{IEEEkeywords} O-RAN, Multi-access Edge Computing, Network Orchestration, Deep Reinforcement Learning, Bayesian Learning\end{IEEEkeywords}

 \section{Introduction}

Open Radio Access Network (O-RAN) is one of the most promising technologies for future RANs \cite{nokia5g, openvran_nec}. In O-RAN, the legacy hardware-based RANs are replaced with softwarized RANs \cite{openvran_nec}. And the Base Station (BS) functions are disaggregated into Radio Unit (RU), Distributed Unit (DU), and Central Unit (CU) \cite{bonati_survey}. The CU and DU can be deployed as Virtual Network Functions (VNFs) and executed through Virtual Machine (VM) instances or lightweight containers across geo-distributed cloud infrastructures \cite{oran_architecture}. It enables flexible deployment and dynamic resource scaling based on users' demands and network conditions, which potentially reduces operational expenses \cite{nokia5g, openvran_nec, bonati_survey, oran_architecture}. Furthermore, another key enabler of 5G+, Multi-access Edge Computing (MEC), brings serverless computing for diverse 5G+ use cases through Function as a Service that can be implemented over virtualized platforms to run the applications. The MEC is expected to serve heterogeneous services, including emerging delay-sensitive applications such as Tactile Internet applications, and hosting them in the same infrastructures as RANs is a way to deliver the services much closer to the users \cite{etsi}.

However, jointly orchestrating the O-RAN and MEC configurations while serving the admitted legacy traffic and heterogeneous MEC demands is non-trivial. Indeed, O-RAN enables a flexible selection of centralization degrees through the functional splits \cite{oran_architecture}. However, such flexibility also creates a problem in finding the optimal split for each BS (which functions are at the DU and CU). The optimal split selection is highly affected by computing resource availability, traffic demands, link capacity, and routing delay between RUs, DUs, and CUs \cite{vran_optimal_murti2}. Each split also induces a different data load over the xHaul links\footnote{The paths connecting a core network (EPC) to CUs, CUs to DUs, and DUs to RUs are backhaul (BH), midhaul (MH), fronthaul (FH), respectively. The integration of these elements is called Crosshaul/xHaul transport network.}, has different constraint requirements, and requires different computing resources. Since the DUs and CUs are virtualized, they need to be allocated a certain amount of virtualized computing resources (e.g., virtual CPU, storage and memory), where their optimal allocation depends on the split selections. They also need to be implemented over geo-distributed cloud platforms (servers), which creates a placement problem of the optimal location to execute each DU/CU. In addition to the delay and resource availability dependence, this placement is affected by the decisions of split selections and allocated resources.

The problem becomes more complex when the RANs also need to accommodate the MEC demands that have diverse service requirements, i.e., an autonomous vehicle application has a different maximum allowable delay compared to 3D gaming, Virtual Reality, etc.
To illustrate, the MEC services can be hosted together with the CUs (as proposed in \cite{etsi}) or even with the DUs (as suggested by experimental studies in \cite{mec_cran_experiment,mec_cran_experiment2}). Such flexibility raises another placement problem in finding the optimal hosting location for each MEC service, either with the DUs or CUs. Hosting the MEC services with the CUs (at a more centralized server) may have processing cost/performance gain, i.e., due to resource pooling and a more powerful computing server, but it induces a higher routing delay \cite{andres_fluidran_joint}. Contrarily, the MEC services can be deployed closer to the users by hosting them with the DUs to reduce the incurred delay \cite{oran_mec_fraga}. When allocating the computing resources and determining the placement locations for these services, we should consider the resource availability, which is affected by the resources, locations, and splits of O-RAN. Clearly, this pairing makes the decisions among O-RAN/MEC configurations intertwined.

Furthermore, the legacy traffic/MEC demands and resource availability might vary over time, suggesting to dynamically reconfigure the O-RAN/MEC system to adjust to the varying conditions. 
However, altering the O-RAN/MEC configurations at runtime may require additional costs or even disrupts the network operations. Such a reconfiguration should be prudently performed by considering long-term consequences. In addition, softwarized RANs have different behavior than legacy RANs, where their configurations often have non-trivial relations, high variance, and dependence on platform and platform load \cite{vranai_journal, edgebol}. Therefore, it becomes impractical to obtain the perfect model of the underlying system and find the optimal configurations over time.

To this end, albeit having minimal modeling assumptions about the underlying system, we aim to solve the above O-RAN/MEC orchestration problem by \textit{dynamically} controlling: \textit{(i)} the split selection for each BS, \textit{(ii)} how much the allocated resources for each DU/CU and MEC service, \textit{(iii)} where to host each DU/CU and MEC service, and \textit{(iv)} how to route the legacy traffic/MEC demands between RUs, DUs, and CUs. Our objective here these decisions are being made to minimize the long-term operating expenses of the network, and at the same time, to maximize the MEC performance criterion.


\subsection{Methodology and Contributions}

We propose and study the joint O-RAN/MEC orchestration problem, where it jointly controls the split selections for the BSs, the resource allocation and placement for each DU/CU/MEC service over geo-distributed platforms, and the routing for each data to the hosting locations. 
Our system model follows the latest proposal of O-RAN architecture with multiple BSs sharing the same computing and link resources. We model the operations as a time-slotted system, where at each time slot, there are arbitrary incoming legacy traffic and MEC demands and resource availability. At each time slot, the control decisions are being selected to minimize the long-term overall operation cost and maximize the MEC performance criterion. This sequential decision-making problem is formulated as Markov decision process (MDP). 

Since obtaining the exact model for the underlying O-RAN/MEC system is non-trivial and it is possibly unknown in practice, our solution framework is developed using a model-free reinforcement learning (RL) approach, where the O-RAN/MEC system is seen as a black-box environment, and we do not make any particular assumptions about the underlying system and state transition probability distribution. However, the resulting formulation renders a semi-continuous state space with multi-dimensional discrete action space, which causes a curse dimensionality issue. To address the dimensionality issue of the state space with discrete action, we adopt a Double Deep Q-network (DDQN)-based approach \cite{ddqn}. Since the action space is also multi-dimensional discrete, the number of estimated outputs for DDQN is expected to grow combinatorially with the number of control decisions (e.g., the number of O-RAN/MEC configurations and BSs). 
%
To tackle such prohibitive complexity, we incorporate action branching \cite{bdq}, an action decomposition method that decomposes the multi-dimension action into sub-actions and utilizes shared decision module followed by neural network branches, with DDQN (BDDQN). This decomposition exhibits a linear growth of the number of estimated outputs while still maintaining the shared decisions. The proposed branching in \cite{bdq} assumes that each sub-action has the same dimensional size, while we adopt it suited to our problem, where each sub-action can have a different size.

However, solving a high-dimensional MDP typically requires numerous trial-and-error interactions. It can become a time-consuming and costly operation, particularly when reconfiguring the virtualized resources (VMs/containers) of the O-RAN/MEC system is expensive and incurs overhead delay. In this case, an efficient exploration-exploitation strategy plays a crucial role. Motivated by the advantages of using a Bayesian neural network \cite{bayes_dqn, bayes_showdown}, we propose a Bayesian framework-based Thompson sampling to encourage data-efficient exploration and improve learning performance. We tailor Gaussian Bayesian Linear Regression (BLR) \cite{gaussian_process} into BDDQN by modifying the output/last layer at each branch to approximate the posterior distribution of the set of Q values. Hence, Bayesian BDDQN not only utilizes estimates of the Q values but also exploits uncertainties over the estimated Q values and employs them to perform Thompson sampling.


Further, we evaluate our proposed approach using collected traces from real demands and a range of network topologies. Following the evaluation results, we proved that our approach is data-efficient, where it significantly converges faster and improves the learning performance by up to 32\% than its non-Bayesian counterpart. Moreover, it offers the cost-saving benefits by 41\% compared to DDPG.
%
We summary our contributions as follows:
\begin{itemize}
	\item We propose and study the joint O-RAN/MEC orchestration problem, where it jointly controls the O-RAN functional splits, the allocated resources and placement locations of O-RAN/MEC services, and the routing for each O-RAN/MEC data flow. This problem is formulated as MDP. 
	\item We propose a novel model-free deep RL framework, Bayesian BDDQN, to solve the formulated MDP. It is constructed from action branching of DDQN (BDDQN) to tackle the multi-dimensional and large action space with linear growth of the neural network outputs. Further, we tailor Bayesian learning into BDDQN by modifying the last layer at each branch to exploit uncertainties to enable data-efficient exploration while also improving the learning performance. It is the first work that tailors Bayesian learning with a branching DDQN algorithm. 
	\item We perform a battery of tests on our approach using O-RAN compliant model and collected measurement traces from real traffic demands. 
\end{itemize}

The rest of this paper is organized as follows. Sec. \ref{sec:relatedworks} discusses our contributions with regards to prior works. In \ref{sec:model}, we discuss the O-RAN/MEC orchestration model and the formulated MDP problem. In Sec \ref{sec:algorithm}, we explain how we design our solution approach, Bayesian BDDQN. The detailed experiment setups and simulation results are discussed in Sec \ref{sec:results}. Finally, we conclude this paper in Sec. \ref{sec:conclusion}.


%
%
%

\section{Related Work} \label{sec:relatedworks}

\textbf{RAN orchestration.} There are some works studied RAN orhestration. For example, by using predetermined models, \cite{dynamic_split_alba} studied altering the functional split at runtime to maximize throughput, to maximize revenue \cite{alba_cost_split}, and to minimize the inter-cell interference and FH utilization \cite{flex5g}. The work in \cite{Bega2018} aimed to maximize the served traffic by efficiently schedule the radio/computing resources. 
Further, other studies proposed model-free approaches such as for orchestrating the radio resource with functional split \cite{Matoussi2020}, managing the interplay between computing and radio resource \cite{vranai_journal}, energy-aware BS \cite{bayes_vran_journal}, green RAN-based functional split selections \cite{Pamuklu2021}, and joint RAN slicing, scheduling and online model training \cite{bonati_coloran}. Possibly, \cite{lofv} and \cite{Murti2023} are the closest RAN orchestration with this paper, where the proposed frameworks jointly control the splits, resources and placement location, and routing for each data flow.
However, none of these work study on the resource sharing between RANs and MEC, although their parameters are highly coupled. 

\textbf{Joint O-RAN/MEC orchestration.} The idea of deploying MEC with O-RAN has been proposed by \cite{etsi, oran_mec_kuklinski} and it is followed by experimental studies of the DU and CU to share their resources with MEC services \cite{mec_cran_experiment,mec_cran_experiment2}. Recent works study how to manage RANs and MEC parameters together. For example, \cite{andres_fluidran_joint} proposed an optimal network design framework to jointly optimize the functional split of RANs and MEC deployment, \cite{oran_mec_fraga} proposed additional degrees of freedom, where the operators can also execute their CU/DU/MEC at servers located in several locations, \cite{soran} expanded the problem in \cite{andres_fluidran_joint} with RAN slicing, where the operators can dynamically reconfigure the isolated slices of RAN and MEC functions together, and \cite{doro_sledge} proposed fast and near-optimal algorithms for networking, storage and computation resources in the joint network-MEC system. 
However, these works relied on fine-tuning models and assumptions, while we adopt model-free approaches.
Using model free approaches, \cite{concordia} developed an ML-based predictor that learns to efficiently share the computing resource between RAN and other workflows, such as edge services. In \cite{edgebol}, the authors proposed a Bayesian online learning for controlling RAN resources and intelligent edge service parameters, aiming to minimize the overall energy cost while satisfying the performance targets. Our orchestration problem differs from these works. In addition to the resource allocation, we consider for the coupling between the functional splits, hosting placement, routing for each service, and the impact of altering the configurations at runtime.

\textbf{Bayesian RL in networking.} One of Bayesian learning approaches for network orchestration that close with our framework is Bayesian contextual bandit. This technique has been applied to minimize the power consumption in virtualized BS \cite{bayes_vran_journal}, to optimize the BS handover \cite{Chuai2019}, to assign CPU of virtualized BS \cite{vranai_journal}, and to jointly control RAN resources with edge AI \cite{edgebol}. However, our orchestration problem requires a full RL formulation while these works only consider the exogenous parameters for the RL state. Perhaps, the closest approach to our framework is action branching method of deep RL for network orchestration as studied in \cite{Murti2023} for O-RAN auto-reconfiguration, \cite{Wei2020} for controlling the network slicing reconfiguration, and \cite{Abdel-Aziz2022} for vehicular networks. However, none of these work adopt Bayesian learning in their deep RL approaches. Here, we tailor an efficient Thompson sampling through Gaussian BLR into the Q and target networks of the deep RL algorithm to enable data-efficient exploration and improve the learning performance.  

\section{System Model and Problem Formulation} \label{sec:model}
%
%
In the latest O-RAN proposals \cite{oran_architecture}, the protocol stacks (or functions) of a BS can be disaggregated into an RU, DU and CU via a functional split. The DU and CU are then hosted as VMs or containers on top of commodity platforms across geo-distributed edge cloud infrastructures.  Similar to O-RAN entities, MEC utilizes a virtualized platform to run the application \cite{etsi}. Hence, O-RAN and MEC can share the same infrastructures. ETSI proposed the MEC deployment on the core network (EPC) or as close as the CU, where the RAN and MEC interaction are performed after the PDCP function (and onwards). Recent studies have experimentally validated and suggested hosting the service co-located with the DU, particularly for delay-sensitive applications, where the interaction can be performed from lower functions through an MEC agent \cite{mec_cran_experiment, mec_cran_experiment2}.
Then, in order to manage resources and interaction of the BSs and MEC, O-RAN has envisioned learning-based orchestration, namely RAN Intelligent Controler (RIC) \cite{bonati_oran}. The controller further is deployed as an xApp in the Non-Real-Time (Non-RT) for closed loop control greater equal than 1 sec and an rApp in Near-Real-Time (Near RT) RIC for 10 ms to 1 sec. Our framework works as an xApp where it enforces a policy at every period of $t = 1, ...., T$. The optimal policy at each time $t$ depends on the state, which is observed at the beginning of each period via O1 interface. 


\begin{table}[t] \centering
	\begin{threeparttable}
		\begin{small}
			\begin{tabular}{@{}lllll@{}}\toprule
				\textbf{}& \textbf{Split Point} & \textbf{Load} &\textbf{Max} & \textbf{Delay Req.}  
				\\ \midrule
				{O1 } &  RRC - PDCP &   $\lambda$ & $ 4$   & $10$ ms         
				\\ \hdashline
				{O2$^*$ } &  PDCP - High RLC &  $\lambda$ & $ 4$  & $10$ ms
				\\ \hdashline
				{O3 } &  High RLC - Low RLC & $\lambda$ & $ 4$    & $10$ ms
				\\ \hdashline
				{O4$^*$ } &       Low RLC - High MAC & $\lambda$ & $ 4$   & $1$ ms         
				\\ \hdashline
				{O5 } & High MAC - Low MAC  & $\lambda$ &  {$ 4$}  & $1$ ms
				\\ \hdashline
				{O6$^*$ } &  Low MAC - High PHY  & $1.02\lambda$+0.5 & {$ 4.13 $}     & $0.25$ ms
				\\ \hdashline
				{O7$^\dagger$ } & High PHY - Low PHY &   {$ 10.1$} &  {$ 10.1$}   & $0.25$ ms
				\\ \hdashline
				{O8$^\dagger$ } &   Low PHY - RF  & {$157.3$}  &{$157.3$}   & $0.25$ ms \\ 
				\bottomrule
			\end{tabular}
			\begin{tablenotes}
				\item Note: $^*$ is applied options for HLS and $^\dagger$ is applied options for LLS. The data load is in Gbps.
			\end{tablenotes}
		\end{small}
	\end{threeparttable}
	\caption{\small The functional split options and their requirements based on 3GPP nomenclature when the traffic demand is $\lambda$ Gbps. The requirements are tailored by following settings: 100 MHz bandwidth, 256 QAM, 32 antenna ports and 8 MIMO layers. The achievable data rate is up to 4 Gbps.}
	\label{table:3gpp_split}
\end{table}

\subsection{Model}

\begin{figure}[t!]
	\centering
	\includegraphics[width=0.48\textwidth]{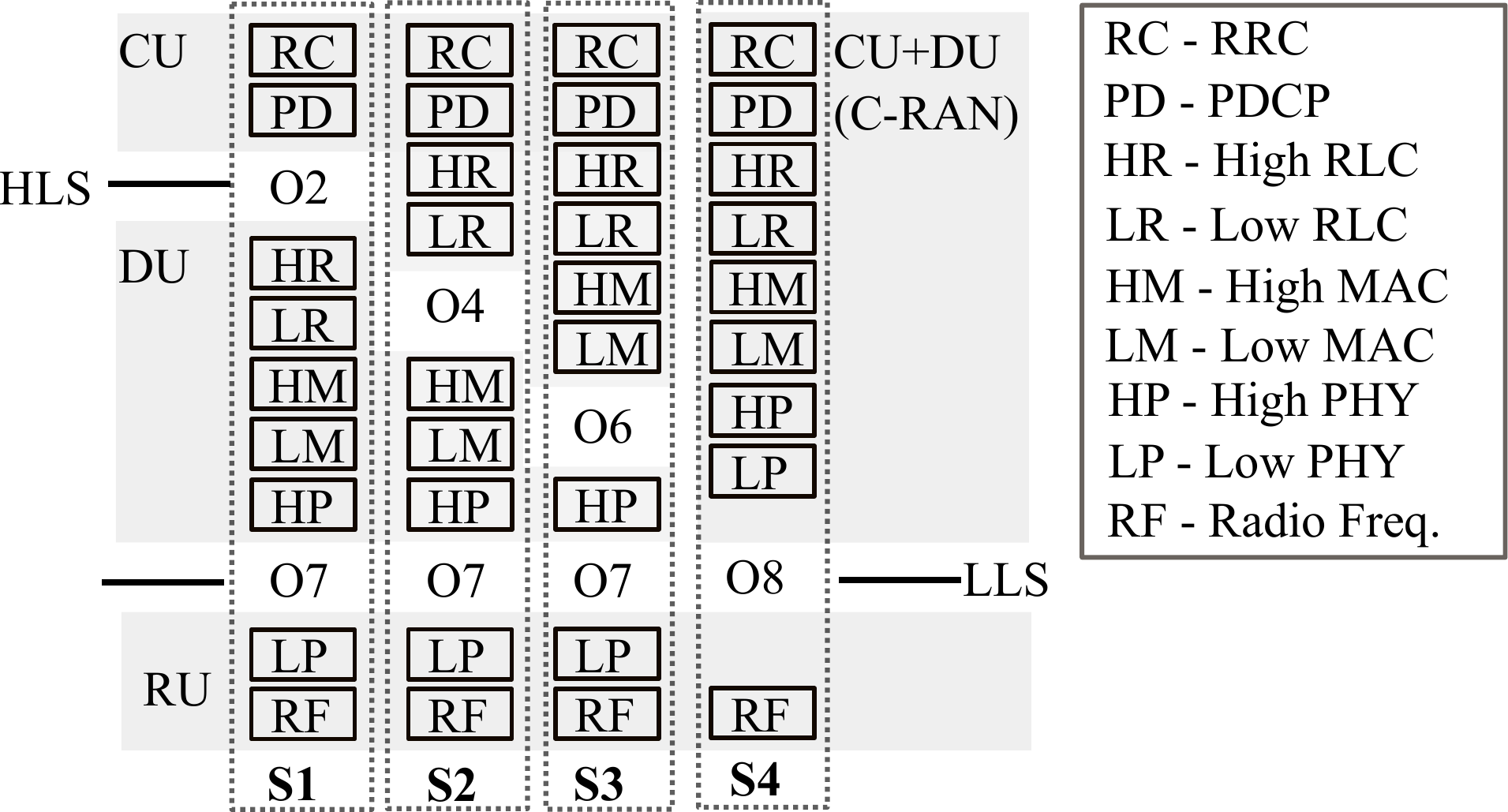}   
	\caption{\small The functional splits applied in our model. S1, S2 and S3 are envisioned by O-RAN proposals, while S4 are legacy C-RAN. }	
	\label{fig:split_option}
\end{figure}

\textbf{Functional Split \& MEC.} Let us consider O-RAN/MEC system with $K$ BSs, where each BS-$k$ can be disaggregated into RU-$k$, DU-$k$ and CU-$k$. 
The DUs and CUs are virtualized components (e.g., VM/container workloads) that can be executed at commodity platforms (e.g., white-box servers), while RUs are radio units. The detailed the functional split nomenclature and their requirements have been defined by 3GPP in \cite{3gpp_rel16}, summarized in Table \ref{table:3gpp_split}.
Following the latest O-RAN proposals, the split between  the DU and RU, called the Low Layer Split (LLS), can implement Option 7.x (O7) and Option 8 (O8). And, the split between the CU and DU, called the High Layer Split (HLS), can employ Option 2 (O2), Option 4 (O4) and Option 6 (O6). Then, we have four selections of the functional splits for our model: \textit{Split 1 (S1)} -- O2 for the HLS and O7 for the LLS; \textit{Split 2 (S2)} -- O4 for the HLS and O7 for the LLS; \textit{Split 3 (S3)} -- O6 for the HLS and O7 for the LLS; and \textit{Split 4 (S4)} -- legacy C-RAN system, which implements Option 8 (O8), i.e., all the BBU are hosted at the servers as an integrated DU/CU and the RF functions are at the RU. We define the possible splits that can be deployed at each BS by the set $\mathcal{V} = \{ \text{S1,S2,S3,S4} \}$, which is illustrated in Fig. \ref{fig:split_option}. And, the MEC services can be hosted together with these functions (e.g., co-located with the CUs or DUs). 

\textbf{Demand.} We focus on the uplink demands as the MEC data typically goes upstream, but it can easily be extended for the downlink. The O-RAN/MEC system must serve the incoming demands from legacy traffic (e.g., mobile broadband) and heterogeneous MEC services. Each different type of service is isolated from others via \textit{hard slicing}, but \textit{soft slicing} (e.g., spatial multiplexing) is applied among same type of services when they are at the same location/server.
We model the incoming demands from the users associated to BS $k$ at time $t$ as follow: \textit{i)} $\lambda_{kc}^{t}$ (Mbps) is the demand from the aggregated request of MEC service type $c \in \mathcal{C}$ that need to be transferred to the MEC hosting location; and \textit{ii)} $\lambda_{k0}^{t}$ (Mbps) is the legacy traffic (type $0$) that must be routed from RU-$k$ to the hosting locations of DU-$k$ and CU-$k$, and then to the EPC/Internet.
All these demands associated with BS-$k$ are denoted by $\lambda^t_k = \big\{ \lambda_{kc}^t :  c \in \mathcal{\tilde{C}} := 0 \cup \mathcal{C} \big\}$, and $\lambda^t = \{ \lambda_k :  k \in \mathcal{K} \}$ defines the set of the demands originated all the BSs.

\begin{figure}[t!]
	\centering
	\includegraphics[width=0.49\textwidth]{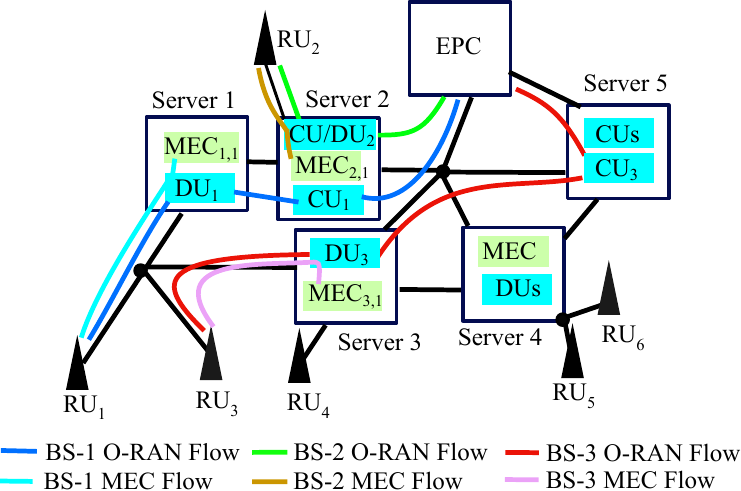}   
	\caption{\small{An example of the joint O-RAN/MEC deployment scenario, where MEC services can be processed at the DU/CU and legacy traffic is further routed to EPC/internet. These diverse services are sharing the same computing and network infrastructures. }}
	\label{fig:network}
	\vspace{-1mm}
\end{figure}

\textbf{Network \& Server.}  We model a packet-based O-RAN/MEC network as a graph of $\mathcal{G} = (\mathcal{I}, \mathcal{E})$, where $\mathcal{I}$ is the set of physical nodes, which includes the subsets: $\mathcal{K} = \{1, ..., K\}$ of RUs, $\mathcal{L} = \{1, ..., L\}$ of hosting platforms (servers), EPC (index 0) and routers; and $\mathcal{E}$ is a set of available links. The DUs are usually hosted at the far-edge servers (close to the RUs) while CUs are at more centralized locations. We define $\mathcal{L}_D  \subseteq  \mathcal{L}$ and $\mathcal{L}_C  \subseteq  \mathcal{L}$ as the sets of candidate servers to host DUs and CUs, respectively. The MEC services can be processed at the same hosting servers (co-located) with DUs and CUs\footnote{The MEC services are originally proposed to be deployed with EPC or CUs (PDCP function and onwards) \cite{andres_fluidran_joint}. However, recent study experimentally validated that they can be deployed at the DUs using an MEC agent \cite{mec_cran_experiment, mec_cran_experiment2}.}, and we define the set of available servers that can process MEC requests with $\mathcal{L}_E \subseteq  (\mathcal{L}_D \cup \mathcal{L}_C) \subseteq \mathcal{L}$.
%
Each server has processing capacity $P_l, \forall l \in \mathcal{L}$. The execution of the DUs, CUs and MEC services needs a certain number of computing resources, where $\mathcal{X}_0$ and $\mathcal{X}_c, c \in \mathcal{C}$ are the set of available configurations (flavors) that determine the amount of the allocated computing resources for CUs/DUs and MEC services, respectively\footnote{We use a different set of flavors between O-RAN and MEC as they may require a different type of resources, i.e., legacy traffic affects CPU resources \cite{vranai_journal} while mobile edge AI needs GPU resources for its processing \cite{edgebol}. }.

Further, each node is connected through link $(i,j) \in \mathcal{E}$, which has a data transfer capacity $c_{ij}$ (Mbps) and delay $d_{ij}$ (secs). Let $p_0 \in \mathcal{P}_k$ and $p_c \in \mathcal{P}_k$ denote the paths to transfer the data flow of legacy traffic and MEC type $c$, respectively, from RU-$k$ to the destination server; and $\mathcal{P}_k$ is the set of all paths connecting RU-$k$ to the servers.
 Then, we have $\mathcal{P} = \cup_{k=1}^K \mathcal{P}_k$ as a set of all paths connecting RUs to servers and consider the data flow is unsplittable. The path $p_0 \in \mathcal{P}_k$ is selected by $p_0 := p^\text{FH} \cup p^\text{MH} \cup p^\text{BH}$. Then, the paths $p^\text{FH} \in \mathcal{P}_k^\text{FH} \subseteq \mathcal{P}_k, p^\text{MH} \in \mathcal{P}_k^\text{MH} \subseteq \mathcal{P}_k, \text{ and } p^\text{BH} \in \mathcal{P}_k^\text{BH} \subseteq \mathcal{P}_k $ are the shortest path at FH, MH and BH among the set of available paths for RU-$k$. The path $p_c \in \mathcal{P}_k$ is selected depending on the MEC hosting locations. When the MEC services are deployed with the DUs, $p_c := p^\text{FH}$; otherwise, $p_c := p^\text{FH} \cup p^\text{MH}$ (co-located with the CUs).
Each of these paths also have a total delay of its links, and we denote the respective path delay as $d_{p_0}, d_{p_c}, d_{p^\text{FH}},  d_{p^\text{MH}}, \text{ and } d_{p^\text{BH}}$.

Finally, Fig. \ref{fig:network} illustrate an example of our system model.

\subsection{Problem Formulation}

We model the O-RAN/MEC orchestration as a time-slotted system. At each time slot $t$, the operator needs to take an action that controls their O-RAN/MEC system configurations to adapt to time-varying legacy/MEC demands and resource availability while respecting constraint requirements. This sequential decision-making problem is formulated as MDP and formalized as follows.

\subsubsection{Action} We introduce a set of control variables  $v^t := \{ v_k^t \in  \mathcal{V} : k \in \mathcal{K}\}$ to activate the functional splits for the BSs at time $t$. This variable determines which functions of the BS to be placed at DU and CU. We define variables $x^t := \{x_{k}^t \in \mathcal{X}_0 : k \in \mathcal{K}\}$ and $y^t := \{y_{k}^t \in \mathcal{X}_0 :  k \in \mathcal{K}\}$ to determine the allocated resources (flavors) for DUs and CUs at the servers. The MEC services can be hosted at the same server with DUs and CUs. And we define the allocated resources for them with $z^t := \{z_{kc}^t \in \mathcal{X}_c : c \in \mathcal{{C}}, k \in \mathcal{K}\}$, where their placement over DUs/CUs is defined by a set of binary variables  $\zeta^t := \{\zeta_{kc}^t \in \{0,1\} : c \in \mathcal{C}, k \in \mathcal{K} \}$, i.e., it is equal to one if the services are co-located with the CUs and otherwise with the DUs.
The function placement of the DUs and CUs over the candidate servers are defined by $\alpha^t := \{\alpha_k^t \in \mathcal{L}_D \subseteq \mathcal{L}: k \in \mathcal{K} \}$ and $\beta^t := \{\beta^t \in \mathcal{L}_C \subseteq \mathcal{L} : k \in \mathcal{K} \}$, respectively. 
Then, the action at time $t$ can be formalized:
\begin{align}\label{eq:action}
	a^t &:= \{ v^t, x^t, y^t, z^t \alpha^t, \beta^t, \zeta^t \} \in \notag
	\\
	& \ \ \mathcal{A} := \{ \mathcal{V} \times \mathcal{L}_D \times \mathcal{L}_C \times \mathcal{X}_0^{2} \times \mathcal{X}_c^{|\mathcal{C}|} \times \mathbb{Z}^{|\mathcal{C}|} \}^{ |\mathcal{K}|},
\end{align}
where the action space $\mathcal{A}$ is a finite set of all pairs control variables associated with all the BSs. Note that the paths $p_0$ and $p_c$ can be directly determined once the action in \eqref{eq:action} is known; hence, we treat the routing for each data flow as part of the environment.

\subsubsection{State} 
The state observation consists of: \textit{(i)} the incoming demands from legacy traffic and MEC demands $\lambda^t$; \textit{(ii)} the last deployed splits of the BSs $v^{t-1}$; \textit{(iii)} the last allocated DU resources $x^{t-1} $ \textit{(iv)}, CU resources $y^{t-1}$ and \textit{(v)} MEC resources $z^{t-1}$; and \textit{(vi)} the last hosting servers/locations for the DUs $\alpha^{t-1}$, \textit{(vii)} CUs $\beta^{t-1}$ and \textit{(viii)} MEC services $\zeta^t$. It gives us information about \textit{time dynamic} of our variable interests: \textit{(i)} the incoming legacy/MEC demands that must to be served; \textit{(ii)} currently active split at each BS; \textit{(iii)} the resource availability for DU, \textit{(iv)} CU and MEC \textit{(v)} that helps to decide increasing/decreasing the allocated resources;  \textit{(vi)} the availability of the servers to host the DUs, \textit{(vii)} CUs and \textit{(viii)} MEC services. Then, we can formalize the state observation of the O-RAN/MEC orchestration at time $t$ as:
\begin{align}\label{eq:state}
	s^t \! &:= \! \{ \lambda^t, v^{t-1}, x^{t-1}, y^{t-1}, \alpha^{t-1}, \beta^{t-1}, \zeta^{t-1} \} \! \in \notag
	\\
	& \ \ \mathcal{S} \! := \! \{ \mathbb{R}^{|\mathcal{\tilde{C}}|} \times \mathcal{V} \times \mathcal{L} \times \mathcal{X}_0^{2} \times \mathcal{X}_c^{|\mathcal{{C}}|}  \times \mathbb{Z}^{|\mathcal{C}|} \}^{|\mathcal{K}|},
\end{align}
where $\mathcal{S}$ is the state space. The first point is an exogenous parameter that does not depend on the action, but provides the contextual information about the users' needs. The other points are the network state, which are highly affected the last selected action.

\subsubsection{Reward \& Learning Objective} 
In this evaluation, the reward is computed from the total network operation cost and MEC performance criterion, which is in this case, we consider the delay cost of elastic MEC services{\footnote{Note that the inelastic MEC services are considered through hard delay constraints and denoted by the set $\mathcal{C}_A \subseteq \mathcal{C}$. Depending on the applications, each service has a different requirement, such arising in Tactile Internet ($\leq$1ms). Then, $\mathcal{C}_B \subseteq \mathcal{C}$ is a set of delay-elastic MEC services, and we focus on delay-sensitive services. However, other performance criteria can be trivially tailored (e.g., throughput).}. The source of monetary costs for network operation are accounted from reserving the computing resources, instantiation/reconfiguration cost, penalty cost due to SLA violation, and routing cost.

The costs for reserving a certain amount of DU/MEC and CU/MEC computing resources can be calculated as:
\begin{align}\label{eq:computing_cost}
	\sum_{k \in \mathcal{K}} f_\text{DM} \big( x_{k}^t + \sum_{c \in \mathcal{C}} (1-\zeta_{kc}^t) z_{kc}^t \big),  
\end{align}
\begin{align}\label{eq:computing_cost2}
	\sum_{k \in \mathcal{K}} f_\text{CM} \big( y_{k}^t + \sum_{c \in \mathcal{C}} \zeta_{kc} z_{kc}^t \big).
\end{align}
The left hand sides in \eqref{eq:computing_cost} and \eqref{eq:computing_cost2} represent the allocated resources for each DU and CU, while the right hand sides represent the allocated resources for MEC, which depend on the MEC placement location, e.g., whether co-located with the DU or CU.
The cost functions $f_\text{DM} (x_k^t) \text{ and } f_\text{CM} (y_k^t)$ charge the allocated DU/MEC and CU/MEC resources into monetary units (\$). 

When allocating the resources, the operator must respect the actual resource utilization. If the allocated resources are less than the actual resource utilization, it can cause some declined/disrupted service demands that can trigger monetary compensation due to SLA violation. And we define this cost:
\begin{align}\label{eq:penalty_underprovisioning}
	\sum_{k \in \mathcal{K} }  f_\text{D} \big( \max (0, \hat{x}_{k}^t - x_{k}^t, \hat{y}_{k}^t - y_{k}^t) + \sum_{c \in \mathcal{C}} \max (0, \hat{z}_{kc}^t - z_{kc}^t)   \big).
\end{align}
where $f_\text{D}(.)$ is the penalty cost function, $\hat{x}_k^t$ is the actual resource utilization of DU-$k$, $\hat{y}_k^t$ is the respective utilization of CU-$k$, and $\hat{z}_{kc}^t$ is for MEC service type $c$ associated with BS-$k$. Note that for simple case, $\hat{x}_k^t$, $\hat{y}_k^t$ and $\hat{z}_{kc}^t$ can be linear relations with their demands. \cite{vran_optimal_murti2}. However, in practice, the actual resource utilization also depends on hosting platform, resource availability, and many unknown factors \cite{vranai_journal}. Hence, we characterize them using the collected measurement traces.
Further, the penalty can be also induced when the enforced action violates the constraint requirements. For instance, the total allocated resources must not exceed their server capacity:
%
%
%
%
\begin{align}\label{eq:penalty_server}
	&\sum_{l \in \mathcal{L}} f_\text{D}  \Big( \max \Big( 0,  \sum_{k \in \mathcal{K}} \big( \mathbbm{1}_{=l} (\alpha_k^t) (x_{k}^t + \sum_{c \in \mathcal{C}} (1-\zeta_{kc}^t)   z_{kc}^t) \notag
	\\
	& \ \ + \mathbbm{1}_{= l} (\beta_k^t) (y_{k}^t + \sum_{c \in \mathcal{C} }   \zeta_{kc}^t  z_{kc}^t) \big) - P_l  \Big).
\end{align}
The deployed configurations also have to respect both O-RAN and MEC service requirements. In O-RAN, the incurred delay at HLS/LLS of each BS has to respect the delay requirement of the selected split:
\begin{align}\label{eq:penalty_delay}
	\sum_{p \in \mathcal{P}_k, k \in \mathcal{K}  } f_\text{D} \big( \max (0, d_{p^\text{FH}} - d_v^L, d_{p^\text{MH}} - d_v^H \big), 
\end{align}
where $ d_v^H$ (secs) and $ d_v^L$ (secs) are the delay requirements of split $v$ at HLS and LLS, respectively,  as seen in Table \ref{table:3gpp_split}. 
Also, each inelastic MEC service has a delay requirement:
\begin{align}\label{eq:penalty_delay_mec}
	\sum_{k \in \mathcal{K}, c \in \mathcal{C}_A  } f_\text{D} \big( \max (0, D^c_k - d^\text{th}_c), 
\end{align}
where $D_{kc}$ and $d^\text{th}_c$ are the total delay and the delay threshold for each inelastic MEC service $c \in \mathcal{C}_A$, respectively. 
In our evaluation, the total delay $D_{kc}$ is calculated from the routing delay $d_{p_c}$ and processing delay. We follow the delay model from an experimental study of OpenFace \cite{andres_fluidran_joint} in C-RAN, which can be calculated:
%
%
\begin{align} \label{eq:delay_mec}
	D_{kc} := \lambda_{kc} d_{p_c} + \delta_1 \big( \lambda_{kc} \frac{\rho_l}{z_{kc}}  \big) + \delta_2 \big(  \frac{\hat{z}_{kc}}{P_l} )^2,
\end{align}	 
where $\rho_l$ is the computational processing capability (Mbps/cycles) of server $l$. The routing delay is calculated depending on the hosting location: $d_{p_c} := (1-\zeta_{kc}^t) d_{p^\text{MH}} + d_{p^\text{FH}}$. And the processing delay is constant at a very low demand, but it increases with the demand for a service that consumes high computing resources.

In order to avoid the above SLA violation (due to insufficient resource allocation and constraint violation) or wasted resources (due to resource overprovisioning), the operator can dynamically instantiate additional resource and reconfigure the O-RAN/MEC system following control variables in \eqref{eq:action}. However, altering the configurations at runtime may require additional costs or even disrupt the network operations. We define the cost that arises for instantiating additional resources:
\begin{align}\label{eq:instantiation}
	& \sum_{k \in \mathcal{K}}  f_\text{I} \big( \max (0, x_{k}^t - x_{k}^{t-1}) + \max (0, y_{k}^t - y_{k}^{t-1})  \notag
	\\
	& \ \ + \sum_{c \in \mathcal{C}} \max (0, z_{kc}^t - z_{kc}^{t-1})  \big).
\end{align}
Then, the reconfiguration cost can arise when the operator decides to alter the BS split or reallocate a new flavor:
\begin{align}\label{eq:reconfiguration}
	\!\!\!\sum_{k \in \mathcal{K}} f_\text{R} \Big( |x_{k}^t - x_{k}^{t-1}| + |y_{k}^t - y_{k}^{t-1}| + \sum_{c \in \mathcal{C}} |y_{kc}^t - y_{kc}^{t-1}| \Big).
\end{align}
This cost also arises when migrating the MEC hosting location between DU and CU:
\begin{align}\label{eq:reconfiguration2}
	\sum_{k \in \mathcal{K}} f_\text{R} \Big( \sum_{c \in \mathcal{C}} z_{kc}^t |\zeta^t_{kc}  -\zeta_{kc}^{t-1}| \Big), 
\end{align}
or migrating the DU/CU resources to a new server location:
\begin{align}\label{eq:reconfiguratio3}
	&\sum_{k \in \mathcal{K}} f_\text{R} \Big(  \big(x_{k}^t + \sum_{c \in \mathcal{C}} (1- \zeta_{kc}^t) z_{kc}^t \big) \mathbbm{1}_{\neq \alpha^{t-1} }(\alpha_k^{t})
	 \notag \\
	 & \ \  + \big(y_{k}^t + \sum_{c \in \mathcal{C}} \zeta_{kc}^t z_{kc}^t \big) \mathbbm{1}_{\neq \beta_k^{t-1} }(\beta_k^{t})  \Big), 
\end{align}
where $f_\text{I}(.) \text{ and } f_\text{R}(.)$ are the instantiation and reconfiguration cost functions. 
When moving the CU/DU/MEC instances into a new hosting location, the whole resources of an instance have to be migrated. A soft migration method can be applied to avoid disrupted operation by creating a duplicate instance that has the same functionality with the old one. Hence, all the new created resources at the new location are accounted for the reconfiguration cost. However, when instantiating/reallocating a new resource at the same server, the raised overhead cost is calculated from the difference between the old and new allocated resource \cite{aztec}.

In addition to the above computing-related resource managements, we also have the routing cost for reserving a bandwidth to transfer the data flow to the destination, which can be denoted: 
\begin{align}\label{eq:routing}
	 \sum_{ k \in \mathcal{K}, p \in \mathcal{P}_k}  f_\text{H} \big(   r_{p^\text{FH}, v} + r_{p^\text{MH}, v} + r_{p^\text{BH}, v} \big)
\end{align}
where $r_{p^\text{FH}, v}, r_{p^\text{MH}, v}, \text{ and } r_{p^\text{BH}, v}$ are the transferred data flow over FH, MH and BH, and $f_\text{H}(.)$ is the routing cost function. This routing cost depends on the selected splits, incurred data flows (see Table \ref{table:3gpp_split}), and hosting locations. 

\textbf{Reward.}  We define the reward with the case where the operator aims to minimize the overall operation cost and maximize MEC performance criterion. And we define this through a scalarized reward as:
\begin{align}\label{eq:reward}
	r(s^t, a^t ) := -J(s^t, a^t) + \eta  B(D(s^t, a^t )).
\end{align}
where $J(s^t, a^t)$ is the overall operation cost, parameter $\eta$ defines the relative importance of the delay cost $D(s^t, a^t ) := \sum_{k \in \mathcal{K}, c \in \mathcal{C}_B} D_{kc} (s^t, a_{kc}^t)$ to the operation cost, and $B$ is a smooth function that models the delay cost associated to elastic MEC services. The overall network cost $J(s^t, a^t)$ is accounted from outputs of the cost functions $f_\text{DM}, f_\text{CM}, f_\text{D}, f_\text{I}, f_\text{R} \text{ and } f_\text{H}$. For simplicity, we assume that these costs functions are proportional with their inputs and introduce the respective coefficients $\kappa_\text{DM}, \kappa_\text{CM}, \kappa_\text{D}, \kappa_\text{I}, \kappa_\text{R} \text{ and } \kappa_\text{H}$ (\$/units) that charge every unit of the input into monetary units (\$), e.g., $ f_\text{DM} (v) := \kappa_\text{DM} v$. Similarly, the smooth function $B(.)$ is also assumed as a linear function that maps the incurred delay into a monetary value (e.g., negative reward).

\textbf{Objective.} The objective of our learning framework is to find an optimal policy that takes a sequence of actions from the action space given a sequence of state observations, which maximize the expected long-term accumulated reward starting at time slot $\tau$, $\mathbb{E}_\pi \sum_{\tau = 0}^{\infty}[\gamma^\tau r^{\tau + t} | \pi]$, as:
\begin{align}\label{eq:policy}
	\pi^*(s) := \arg\max \mathbb{E}_\pi \sum_{\tau = 0}^{\infty}[\gamma^\tau r^{\tau + t} | \pi],
\end{align}
where the discount $\gamma$ is set to $\gamma = 1$ during the online operation; otherwise, $\gamma \in (0,1]$.


\section{Bayesian BDDQN Algorithm} \label{sec:algorithm}
We design our solution framework following a model-free RL paradigm, which treats the O-RAN/MEC system as a black-box environment and imposes no assumptions about the system state and state transition probability distribution. However, the formulated MDP raises challenging dimensionality issues because the state space is semi-continuous and the action space is multi-dimensional discrete. To address such a challenging RL problem, we proposed a novel agent called Bayesian BDDQN.  
It is built from DDQN \cite{ddqn} to address an RL problem that has a large state space with discrete action space. Since the discrete action space is multi-dimensional that consists of multiple degrees of freedom, it renders a combinatorial growth of the number of possible actions that DDQN needs to estimate. Hence, we incorporate an action decomposition method through action branching \cite{bdq} into DDQN (BDDQN) to reduce the complexity into a linear increase. Then, we employ an efficient exploration-exploitation strategy using a Thompson sampling by tailoring a Bayesian framework into BDDQN; hence, it not only exploits the estimated Q functions but also utilizes their uncertainties \cite{bayes_showdown,bayes_dqn}. This strategy becomes essential, particularly when an efficient pre-training model may not be available and performing trial-and-error interactions with the environment becomes time-consuming and costly.   
The detailed design of our proposed Bayesian BDDQN is discussed as follows.

\subsection{DDQN} 
Let us define the optimal action-value function (Q function) $Q^*(s,a)$ as the maximum expected reward after observing some sequences $s$, then following some policies $\pi$ and taking some actions $a$: $Q^*(s,a) := \max_\pi \mathbb{E} [ \sum_{\tau}^{\infty} \gamma r^{\tau+t} | s^t = s, a^t = a]$. We can find the optimal policy $\pi^* := \mathbb{E}_{s \sim} \mathcal{E}  [r + \gamma \max_{a'} Q(s', a') | s',a' ]$ if the the optimal value $Q(s', a')$ given the sequences at the next time slot $s'$ for all the possible actions $a'$ is known. Since finding the optimal Q function via the value iteration method is impractical, the Q function can be estimated through a function approximator such as a neural network \cite{dqn_mnih1}. The estimated Q function parameterized by a neural network (Q-network) with weights $\theta$ can be represented as: $Q(s,a; \theta) \approx Q(s,a)$ and trained by minimizing the loss function:
\begin{align}\label{eq:loss_dqn}
	L(\theta) := \mathbb{E}_{s, a,r, s' \sim \mathcal{D}} \big[ u - Q(s, a; \theta) \big],
\end{align}
where $u$ is the Temporal Difference (TD) target and the transition $\{s,a,r,s'\}$ is collected by a random sampling from the stored experience $\mathcal{D}$. 
The TD target of DQN \cite{dqn_mnih1} is represented by:
\begin{align}\label{eq:target_dqn}
	u^{\text{DQN}} := \mathbb{E}_{s' \sim \mathcal{S}} [r + \gamma \max \tilde{Q}(s', a; \tilde{\theta})],
\end{align}
where $\tilde{Q}(s', a'; \tilde{\theta})$ is the target network parameterized by weights $\tilde{\theta}$. The TD target in DQN is frequently overestimated in relation to the actual Q-function. Thus, this overestimation issue is addressed by using Double DQN (DDQN) \cite{ddqn}, which modifies the TD target into: 
\begin{align}\label{eq:target_ddqn}
	u^{\text{DDQN}} := \mathbb{E}_{s' \sim \mathcal{S}} [r + \gamma \tilde{Q}(s', \underset{a'}{\arg\max} Q(s', a'; \theta); \tilde{\theta})].
\end{align} 
%
%

%
%
\subsection{BDDQN}
\begin{figure*}[t!]
	\centering
	\begin{subfigure}[t]{.48\textwidth}
		\includegraphics[width=\textwidth]{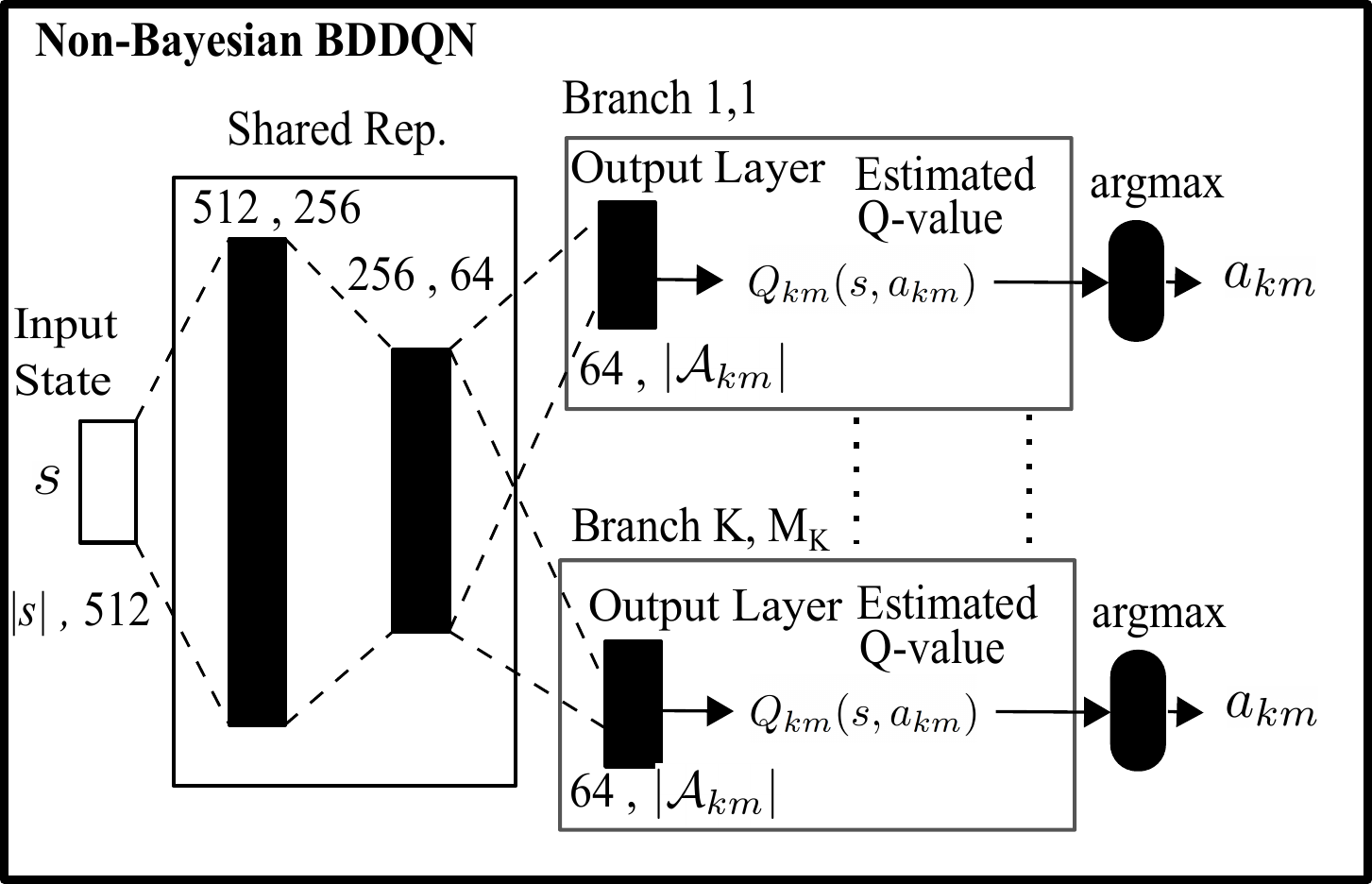}
		\caption{Non-Bayesian BDDQN}
		\label{fig:bddqn}
	\end{subfigure} 
	\hfill
	\begin{subfigure}[t]{.495\textwidth}
		\includegraphics[width=\textwidth]{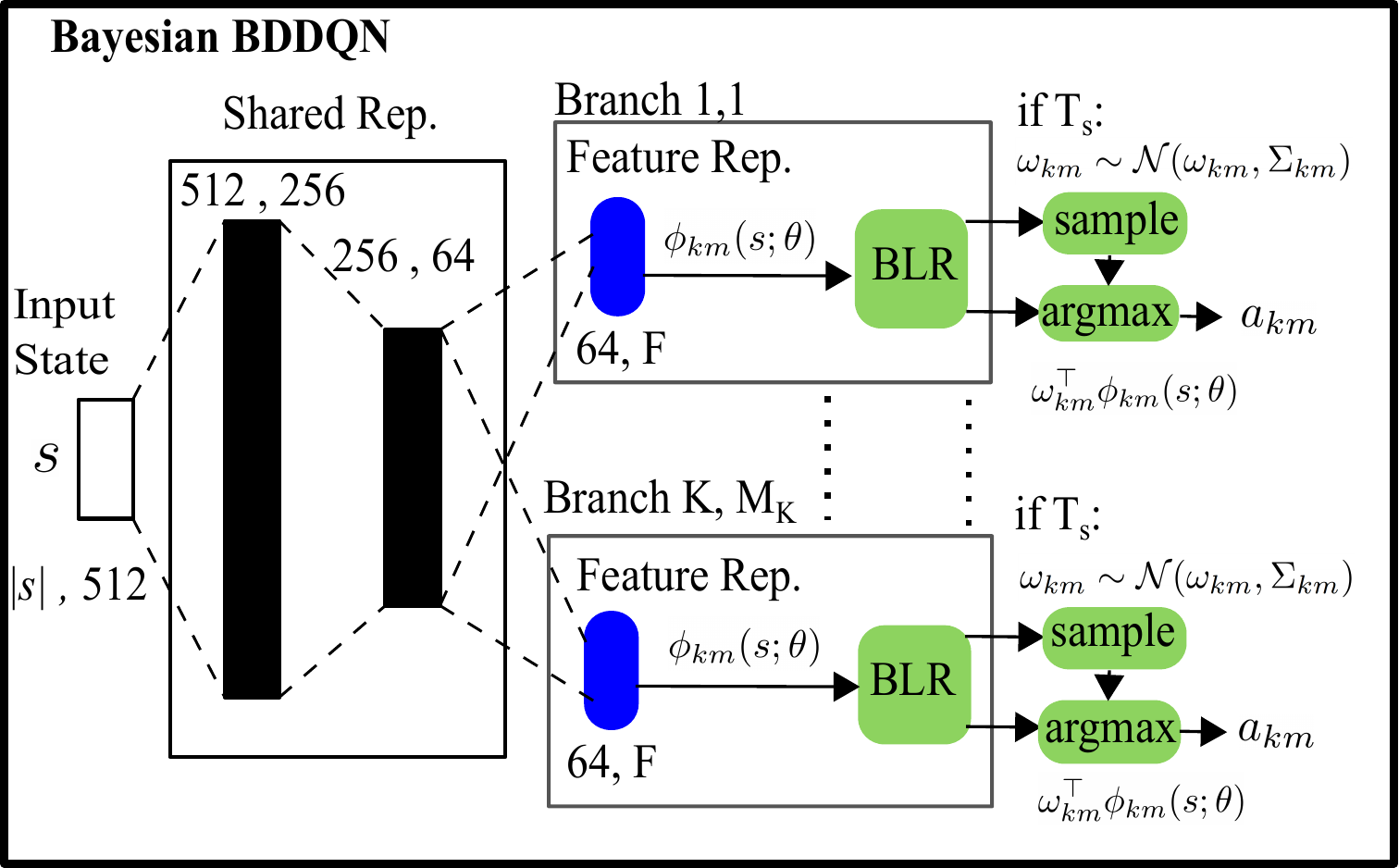}
		\caption{Bayesian BDDQN}
		\label{fig:bayes_bddqn}
	\end{subfigure}     
	\caption{\small The Q network of (a) non-Bayesian BDDQN and (b) Bayesian BDDQN. Bayesian BDDQN removes the last/output (e.g., Linear) layer and replaces it with a feature representation layer at each branch (marked with blue color). Bayesian BDDQN also adds mechanisms for BLR and Thomson sampling after the feature representation layer (marked with green color).  }
	\label{fig:q_network}
\end{figure*}

Although DDQN can effectively address many RL applications with a large state space with discrete action \cite{ddqn}, it does not intend to handle a multi-dimensional discrete action space, such as arising in our problem. It renders a combinatorial growth in the number of estimated Q values with the increase of control decisions (e.g., O-RAN/MEC configurations and BSs).  
We discuss how to adopt an action decomposition method through action branching \cite{bdq} with DDQN (BDDQN) to convert such a combinatorial growth into a linear increase.

We define $\mathcal{M}_k := \{ v_k^t, x_k^t, y_k^t, z_{kc}^t \alpha_k^t, \beta_k^t, \zeta_{kc}^t : c \in \mathcal{C} \} $ as the set of all control variables from the action set defined in \eqref{eq:action} that are associated with each BS-$k$; and $M_k = |\mathcal{M}_k|$. Then, we can decompose the action $a$ into sub-actions $a_{km}, \forall m \in \mathcal{M}_k, \forall k \in \mathcal{K}$, where each sub-action represent each control variable at every BS\footnote{From this point, we use the term sub-action to refer each control variable.}. Hence, the action in \eqref{eq:action} can also be represented by $a := \{a_{km} :
m \in \mathcal{M}_k, k \in \mathcal{K}\}$. Each of sub-actions also takes values from a finite set of the sub-action space $\mathcal{A}_{km} \subseteq \mathcal{A}$ that describes the $m$-th control space associated with BS-$k$.
As the RL problem has $M_k$ sub-actions at every BS, the number of possible actions to be estimated when directly applying DDQN becomes $\prod_{k=1}^{K} \prod_{c=1}^{M_k} |\mathcal{A}_{km}|$. By adopting an action decomposition method such as action branching, the estimated Q values can be reduced into $\sum_{k = 1}^K \sum_{c = 1}^{M_k} |\mathcal{A}_{km}|$. 
The initial method outlined in \cite{bdq} has effectively addressed applications with a discretized continuous action space, but its effectiveness has not yet been proven in scenarios where the action space is inherently multi-dimensional. Additionally, they still assumed that every sub-action space has the same dimensional size. Hence, we can not directly adopt it to our problem. We describe how to tailor action branching to DDQN suited for our RL problem.

We use the common state $s$ (defined in \eqref{eq:state}).  Then, the Q value corresponds to sub-action $a_{km}$ at common state $s$ can be denoted as $Q_{km} (s, a_{km})$ and estimated following the Q network in Fig. \ref{fig:bddqn}. 
%
%
The TD target of BDDQN is set as a single global learning target (e.g., $u_{km} := u, \forall k \in \mathcal{K}, \forall m \in \mathcal{M}$) and similar to the DDQN TD target in \eqref{eq:target_ddqn}, but we average all the dimensions of the sub action:
\begin{align}\label{eq:target_bdq}
	u \!:=\! r \!+\! \gamma \frac{1}{K} \sum_{k=1}^K \! \frac{1}{M_k} \sum_{m=1}^{M_k} \! \tilde{Q}_{km} \big( s', \underset{a_{km} \in \mathcal{A}_{km}}{ \! \arg\max} \! Q_{km}(s', a_{km}) \big),
\end{align}
where $\tilde{Q}_{km}$ is the target network. Then, the Q network is trained to minimize the following loss function:
\begin{align}\label{eq:loss_bdq}
	L(\theta) \! := \!  \mathbb{E}_{s, a, r, s' \sim \mathcal{D}} \big[ \frac{1}{K} \sum_{k = 1}^K \frac{1}{M_k} \sum_{m=1}^{M_k} [u \!-\! Q_{km}(s, a_{km}; \theta) ]^2 \big].
\end{align}

\textbf{BDDQN Architecture.} Fig. \ref{fig:bddqn} shows the Q-network architecture of DDQN. BDDQN is built from an input layer, a shared representation segment with several hidden layers, and neural network branches. 
The input layer is constructed from a Linear layer with ReLU activation function and has an input size $|s|$ to receive the common state observation $s$. 
Then, we build the shared representation segment using two fully connected Linear layers (each layer with ReLU activation). 
The outputs of this shared segment become the input of neural network branches, where each branch aims to estimate the Q-value $Q_{km}(s, a_{km})$. And we simply apply a Linear layer that has an output size of $|\mathcal{A}_{km}|$.
Since BDDQN does not take uncertainties of the estimated Q values into account (e.g., it only computes a point of estimates), it adopts an $\epsilon$-greedy strategy to select the action, i.e., $a \!:=\! \big[ \underset{a_{k1}}{\arg\max  Q_{k1} (s, a_{k1}) }, ...,  \underset{a_{KM_K}}{\arg\max  Q_{KM_K} (s, a_{K,M_K}) } \big]$ by probability $1- \epsilon$ and otherwise, to select randomly.

\subsection{A Bayesian Framework for BDDQN}
Albeit BDDQN is suited to address a large state space with a multi-dimensional action space, one challenging issue rises in our problem is to perform an effective exploration-exploitation strategy. Note that our agent tries to control virtualized resources (e.g., VMs), where trial-and-error interactions with environment can be a time-consuming and costly. And, the pre-training models also may not be always accurate or available. Therefore, we adopt an efficient Thomson sampling-based method under a Bayesian framework for solving our high dimensional RL problem. The idea is to remove the last Linear layer at each branch and replace it with the feature representation layer, where we deploy BLR on top of it.



In BDDQN, the Q and target networks are constructed following a deep neural network architecture illustrated in Fig \ref{fig:bddqn}. It utilizes a Linear layer for the last (output) layer at each branch and the agent learns the Q function through empirical estimates of the regression problem in \eqref{eq:loss_bdq}. Therefore, the Q function (corresponding to possible sub-actions of each branch) can be represented as a linear transformation of features as: $Q_{km}(s,a_{km}) := \phi_{km}(s;\theta)^\top \omega_{a_{km}}$, where $\phi_{km}(s;\theta) \in \mathbb{R}^{d}$ is the feature representation of the Q network and $\omega_{a_{km}} \in \mathbb{R}^d, \forall a_{km} \in \mathcal{A}_{km}$ is the parameters of the last linear layer for each possible sub-action. Since The target network follows the same architecture with the Q network, we can redefine \eqref{eq:target_bdq} by:
\begin{align}\label{eq:target_newbdq}
	u \!:=\! r \!+\! \gamma \frac{1}{K} \sum_{k=1}^K \! \frac{1}{M_k} \sum_{m=1}^{M_k} \! \tilde{\phi}_{km}(s';\tilde{\theta})^\top \tilde{\omega}_{\hat{a}_{km}},
\end{align}
 where  $\tilde{\phi}_{km}(s;\tilde{\theta}^\top) $ and $\tilde{\omega}_{\hat{a}_{km}}$ are the feature representation and last layer parameters for target network; and we have
 \begin{align}\label{eq:a_hat}
\hat{a}_{km} := \arg\max_{a_{km}} \phi_{km}(s';\theta)^\top \omega_{a_{km}}.
 \end{align}
Clearly, the regression in \eqref{eq:loss_bdq} results in a linear regression problem of the last layers. Instead of solving this regression directly (e.g., using a point of estimate), Bayesian BDDQN uses Gaussian BLR, which renders the approximated posterior on the weight parameters $\omega_{a_{km}}$ and eventually the Q function.

Let consider the prior and likelihood choices conjugates to each other, i.e., the posterior resulting from the Bayesian updating process is in the same parametric family as the prior (e.g., Gaussian distribution). Hence, the posterior update can be computationally tractable since there exists a closed form. Given the experience replay buffer $\mathcal{D} = \{s^\tau, a^\tau, u^\tau \}$, we build disjoint datasets: $ \mathcal{D} = \cup_{k=1}^K \cup_{m=1}^{M_k} \cup_{a_{km}=1}^{|\mathcal{A}_{km}|} \mathcal{D}_{a_{km}} $, where $\mathcal{D}_{a_{km}}$ is the datasets correspond to sub-action $a_{km}$. Then, our interest is to approximate the posterior distribution of $\omega_{a_{km}}$ and correspondingly $Q_{km}(s,a_{km})$: $\mathbb{P}(\omega_{a_{km}}, \mathcal{D}_{a_{km}})$ and $\mathbb{P}(Q_{km}(s,a_{km}), \mathcal{D}_{a_{km}})$. Following BLR, for each sub-action $a_{km}$ and the corresponding dataset $\mathcal{D}_{a_{km}}$, we construct $\textbf{u}_{a_{km}} \in \mathbb{R}^{|\mathcal{D}_{a_{km}}|}$ as the concatenation of target values in set $\mathcal{D}_{a_{km}}$ and a matrix $\Phi_{a_{km}}$ from a concatenation of feature column vectors corresponds to each branch as $\Phi_{a_{km}} = \{\phi(s_i) \}_{i=1}^{|\mathcal{D}_{a_{km}}|} \in \mathbb{R}^{d \times |\mathcal{D}_{a_{km}}| }$. Then, the posterior distribution of $\omega_{a_{km}}$ can be represented as:
\begin{align}\label{eq:posterior_wa}
	\omega_{a_{km}} \sim \mathcal{N} (\mu_{a_{km}}, \Sigma_{a_{km}}),
\end{align}
and the mean and variance can be computed as:
\begin{align}\label{eq:mean_wa}
	\mu_{a_{km}} = \frac{1}{\sigma^2_\epsilon} \Sigma_{a_{km}} \Phi_{a_{km}} \mathbf{u}_{a_{km}}
\end{align}
\begin{align}\label{eq:var_wa}
	\Sigma_{a_{km}} = \Big( \frac{1}{\sigma^2_\epsilon} \Phi_{a_{km}} \Phi_{a_{km}}^\top  + \frac{1}{\sigma} \mathbb{I} \Big)^{-1}
\end{align}
where $\mathbb{I}$ is identity matrix and $\sigma_{\epsilon}$ is the standard deviation of bias. 
By utilizing \eqref{eq:posterior_wa}, we can have $Q_{km}(s, a_{km}) | \mathcal{D}_{a_{km}} := \omega_{{a}_{km}}^\top \phi_{a_{km}}(s)$. 

Let $\omega_{km} = \{ \omega_{{a}_{km}}: a_{km} \in \mathcal{A}_{km} \}$, $\tilde{\omega}_{km} = \{ \tilde{\omega}_{{a}_{km}}: a_{km} \in \mathcal{A}_{km} \}$, $\mu_{km} = \{ \mu_{{a}_{km}}: a_{km} \in \mathcal{A}_{km} \}$, and $\Sigma_{km} = \{ \Sigma_{{a}_{km}}: a_{km} \in \mathcal{A}_{km} \}$. We perform Thompson sampling for the exploration-exploitation strategy by adopting \eqref{eq:posterior_wa}. At every time slot, each sub-action can be selected by:
\begin{align}\label{eq:bay_subactions}
	\! \! a_{km} := \arg \max_{a_{km}} \omega_{{km}}^\top \phi_{km} (s;\theta),
\end{align}
and $a := \{ a_{km} : k \in \mathcal{K}, m \in \mathcal{M}_k  \}$. 
 Then, the objective of our agent is to learn the Q network presented in Fig. \ref{fig:bayes_bddqn} by minimizing the loss function:
\begin{align}\label{eq:loss_bayes}
	L(\theta) \! := \!  \mathbb{E}_{s, a, r, s' \sim \mathcal{D}} \big[ \frac{1}{K} \sum_{k = 1}^K \frac{1}{M_k} \sum_{m=1}^{M_k} [u \!-\!  \omega_{{km}}^\top \phi_{km}(s, \theta) ]^2 \big].
\end{align}
The TD target $u$ in \eqref{eq:loss_bayes} above can be computed as:
\begin{align}\label{eq:target_bayes}
	u  := &r \!+\! \gamma \frac{1}{K} \sum_{k=1}^K \! \frac{1}{M_k} \sum_{m=1}^{M_k} \! \tilde{\omega}_{\hat{a}_{km}}^\top \tilde{\phi}_{km}(s';\tilde{\theta}), \text{ where}
	\\
	& \hat{a}_{km} :=  \arg \max_{a_{km}} \omega_{km}^\top   \phi_{km} (s';\theta)
\end{align}
%
%
%
%
%
%
\begin{algorithm}[t!]  
	\caption{Bayesian BDDQN Algorithm}
	\label{alg:bayes}
	\SetAlgoLined
	\DontPrintSemicolon
	\textbf{Initialize:}  $\phi(s; \theta)$ using random or pretraining weights $\theta$ and clone it as $\tilde{\phi}(s; \tilde{\theta}) \leftarrow \phi(s; \theta)$. \\
	\textbf{Initialize:} Replay memory $\mathcal{D}$ with a fixed buffer size. \\
	\textbf{Initialize:} $\omega_{km}, \tilde{\omega}_{km}, \mu_{km}, $ and $ \Sigma_{km}$, $\forall k \in \mathcal{K}, \forall m \in \mathcal{M}_k$, with Gaussian distribution. \\
	Set $count=0$ and update period $T_p, T_s, T_g$.  \\
	\For{Each episode $e=1.., E$}{
		Reset state of all the BSs ${s}^1 \!:=\! \{ \lambda^1, i^{0}, x^{0}, y^{0}, z^{0}, \zeta^{0}\}$. \\
		\For{Each time slot $t=1...,T$}{
			 \If { $ count \mod T_p = 0$}{
			 	Update $\mu_{km}$ and $ \Sigma_{km},$ $\forall k \in \mathcal{K}, \forall m \in \mathcal{M}_k$ with \eqref{eq:mean_wa}-\eqref{eq:var_wa}	
			 }
			Select the action $a^t := \{a^t_{km} : m \in \mathcal{M}_k, k \in \mathcal{K}\}$ using \eqref{eq:bay_subactions}. \\
			Determine the routing $p_c \in \mathcal{P}_k, \forall c \in  \mathcal{\tilde{C}}, \forall k \in \mathcal{K}$ using information obtained from $a^t$\\
			Enforce $a^t$ and $p_c \in \mathcal{\tilde{C}}$ to all the BSs and compute the costs $J(s,a)$ and $D(s,a)$. \\
			Collect the reward $r(s,a)$ based on \eqref{eq:reward}. \\
			Set $s^{t+1} \leftarrow s^t$. \\
			Store the experience $\mathcal{D} \leftarrow \big\{s^t, a^t, r^t, s^{t+1}  \big\}$. \\
			Sample a minibatch of experiences from $\mathcal{D} $. \\
			 Set $u^t :=$ 
			 $\begin{cases}
			 	r(s^t, a^t); \ \ \text{for terminal} \ s^{j+1}
			 	\\
			 	\text{Compute Eq. \eqref{eq:target_bayes}.}  
			 \end{cases}$ \\
			Perform a gradient descent method to the loss function $L(\theta)$ in \eqref{eq:loss_bayes} w.r.t $\theta$. \\
			\If { $ count \mod T_g = 0$}{
				Update target network by $\tilde{\phi}(s;\tilde{\theta}) \leftarrow \phi(s; \theta)$ and $\tilde{\omega}_{km} \leftarrow \mu_{km}, \forall k \in \mathcal{K}, \forall m \in \mathcal{M}_k$.	
			}		
			\If { $ count \mod T_s = 0$}{
				Sample $\omega_{{km}} \sim \mathcal{N}({\omega}_{km}, \Sigma_{km}),\forall k \in \mathcal{K}, \forall m \in \mathcal{M}_k$.				
			}
		$count = count +1 $.
		}
	}
\end{algorithm}
Finally, Algorithm \ref{alg:bayes} summarizes the learning process of Bayesian BDDQN.



\section{Results and Discussion} \label{sec:results}
%
\subsection{Experiment Setup}
%
%

Our simulations rely on a Milan-based MEC topology (N1) \cite{milan_topology} and real traffic demands \cite{milan_datasets}. We also use a synthetic topology (N2) generated using Waxman algorithm \cite{waxman} with parameters of link probability (0.5) and edge length control (0.1). Each BS in N1 and N2 serves three categories of slicing: legacy traffic (e.g., mobile broadband), delay-elastic MEC services (e.g., massive machine-type communications) and delay-inelastic MEC services (e.g., ultra-reliable low latency communications). Since the difficulty capturing the actual computing behavior of the traffic in \cite{milan_datasets} in a tracable model, we utilize a deep neural network model that maps the traffic demands into the actual computing utilization following \cite{Murti2023}\footnote{The model is trained using real collected measurements from two different platforms (e.g., Platform A and B) that run srsRAN \cite{srslte}. Then, the architecture is constructed from an input, an output and three hidden layers. The size of hidden layers are 128, 64 and 16. The model is trained using Adam optimizer over 200 epochs. It uses mini-batches with size of 128 and MSE loss function. The detailed experiments are available in \cite{Murti2023}.}. We use the term Reference Core (RC) to define a computing unit, i.e., 1 RC translates to 1 virtual CPU unit. Among the O-RAN functions, LP, HP, LM, HM, LR, HR and PD yield 48\%, 17\%, 7\%, 7\%, 0.5\%, 0.5\%, 10\%, 10\% of the total BBU computing utilization, respectively, c.f. \cite{andres_fluidran_joint,placeran}. A single cluster of O-RAN/MEC system in N1 and N2 consists of 1 EPC, 6 available servers (4 servers for DUs/MEC and 2 servers for CUs/MEC), and 4 RUs (default), where the routers are co-located with each node. N1 and N2 have per link delay latency between 0 and 0.1 ms, capacity between 30 Gbps and 160 Gbps, and link weights between 0 to 0.1. We set the computing capacity of each server as $H_l = 20 \text{ RCs}, \forall l \in \mathcal{L}_D \subseteq \mathcal{L}$ and $H_l = 100 \text{ RCs}, \forall l \in \mathcal{L}_C \subseteq \mathcal{L}$. The computational processing at each server to process the MEC services is assumed $\rho_l = 1$ /Mbps/cycles, and we set MEC delay parameters $\delta_1 =: \delta_2 = 1$.
For simplicity, we set the available flavors homogeneously for each service with $|\mathcal{X}_c| := |\mathcal{X}_0| := 16$ that translates into $\{ 0,1, .... 13,14, 15\}$ RCs. Then, we define two different environments for our evaluation, where we utilize Platform A with N1 (OM1) and Platform B with (OM2). Unless otherwise stated, we use OM1 as the default environment.

Further, we set the default coefficient fees with $ \kappa_\text{DM} := 0.25, \kappa_\text{D} := 5, \kappa_\text{R} := \kappa_\text{I} := 0.05$, and $\kappa_\text{H} := 1$. Since executing the BS functions or MEC services at a more centralized computing platform can gain central processing benefits, we set $\kappa_\text{CM} := 0.5 \kappa_\text{DM}$ (see \cite[Fig. 6a]{complexity_cran} with $\approx 10$ BSs). The detailed Q-network of Bayesian BDDQN is presented in Fig. \ref{fig:bayes_bddqn}. In total, there are $\sum_{k = 1}^K M_k$ branches, and each branch has a representation layer with the size of $F = 128$ and an output with size of $|\mathcal{A}_{km}|$. The batch size and replay buffer capacity are 128 and $10^6$, respectively. We set the learning rate of Adam optimizer \cite{adam_optim} with $10^{-4}$. The exploration-exploitation strategy is based on Thompson sampling (under a Bayesian BDDQN framework). The time horizon for a single episode is 144 time slots. The agent updates its posterior distribution at every $T_p = 1440$ time slots (10 episodes), the target network at every $T_g = 1440$ time slots (10 episodes), the policy by re-sampling its posterior distribution at every $T_s = 144$ time slots (1 episode). 

\subsection{Numerical Evaluation}
We compare our proposed solution, Bayesian BDDQN, to its non-Bayesian version and a state-of the art RL approach, DDPG. In DDPG, we relax the discrete action defined in \eqref{eq:action} into a continuous action. Then, we discretize the selected action by estimating the output of DDPG into the nearest discrete value. Since the output must be positive, we use a Sigmoid function at the output layer.

\begin{figure}[t!]
	\centering
	\includegraphics[width=0.495\textwidth]{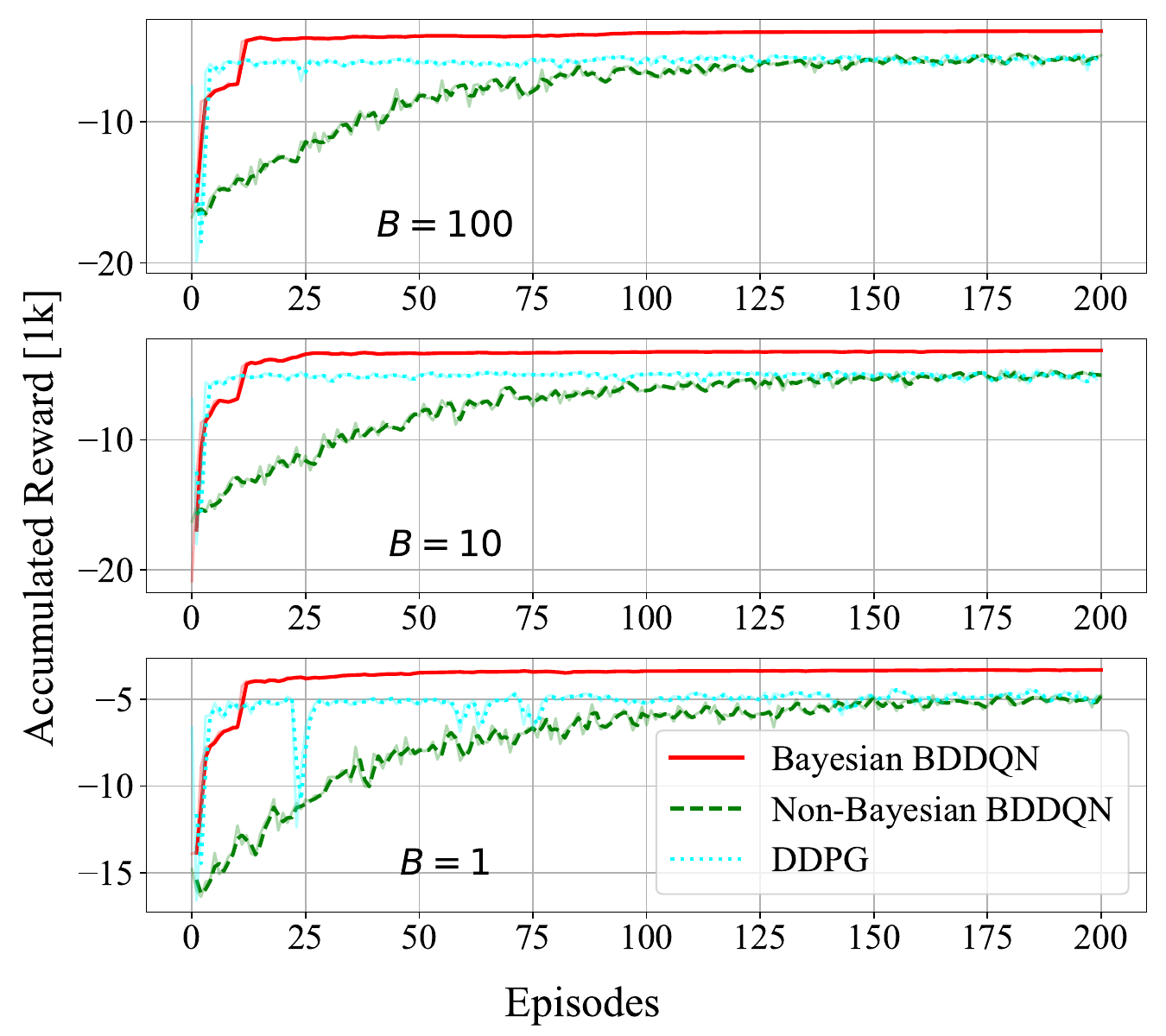}   
	\caption{\small On the comparison of Bayesian BDDQN with other benchmarks over various delay coefficient fees of elastic MEC services.}	
	\label{fig:reward_nopretrain}
\end{figure}

\subsubsection{Delay coefficient fee} Fig. \ref{fig:reward_nopretrain} illustrates the learning performance of Bayesian BDDQN compared to its non-Bayesian version and DDPG over various delay coefficients of the elastic-delay services when the pre-training model is not available. Albeit having different delay coefficients, it shows that Bayesian BDDQN successfully learns the optimal policy and converges to the best policy that the agent can learn. Moreover, it shows that Bayesian BDDQN can obtain a higher return reward than other approaches. Our findings reveal that the episodic reward of BDDQN can be improved by using a Bayesian approach (Bayesian BDDQN) up to 31.25\%, 32.76\%, and 29.76\% when $B=100$, $B=10$, and $B=1$, respectively. Moreover, compared to DDPG, the episodic reward of Bayesian BDDQN is significantly higher by 31.12\%, 31.71\%, and 25.07\% when $B=100$, $B=10$, and $B=1$, respectively. It means that our approach offers a more cost-saving compared to both benchmarks.
%

Furthermore, by adopting Thompson sampling with a Bayesian framework, our approach can also significantly expedite the learning convergence compared to its non-Bayesian version. Fig. \ref{fig:reward_nopretrain} indicates that our approach is data-efficient, which can converge within less than 12 episodes, precisely after the posterior distribution at each branch is updated. On the contrary, the non-Bayesian needs around 125 episodes to converge. Although the convergence speed of DDPG can be as fast as Bayesian BDDQN, DDPG is still underperformed, offering a considerable lower episodic reward. DDPG utilizes Ornstein–Uhlenbeck (OU) noise for the exploration-exploitation strategy, originally used for continuous action spaces, and its reward performance is degraded when the action space is discrete such as arising in our problem. 
\textit{The above evaluations highlight that Bayesian BDDQN is data-efficient (i.e., offering fast learning convergence) and, at the same time, obtains the highest episodic reward (hence, the most cost-efficient) over all the delay coefficient settings.}

\begin{figure}[t!]
	\centering
	\includegraphics[width=0.48\textwidth]{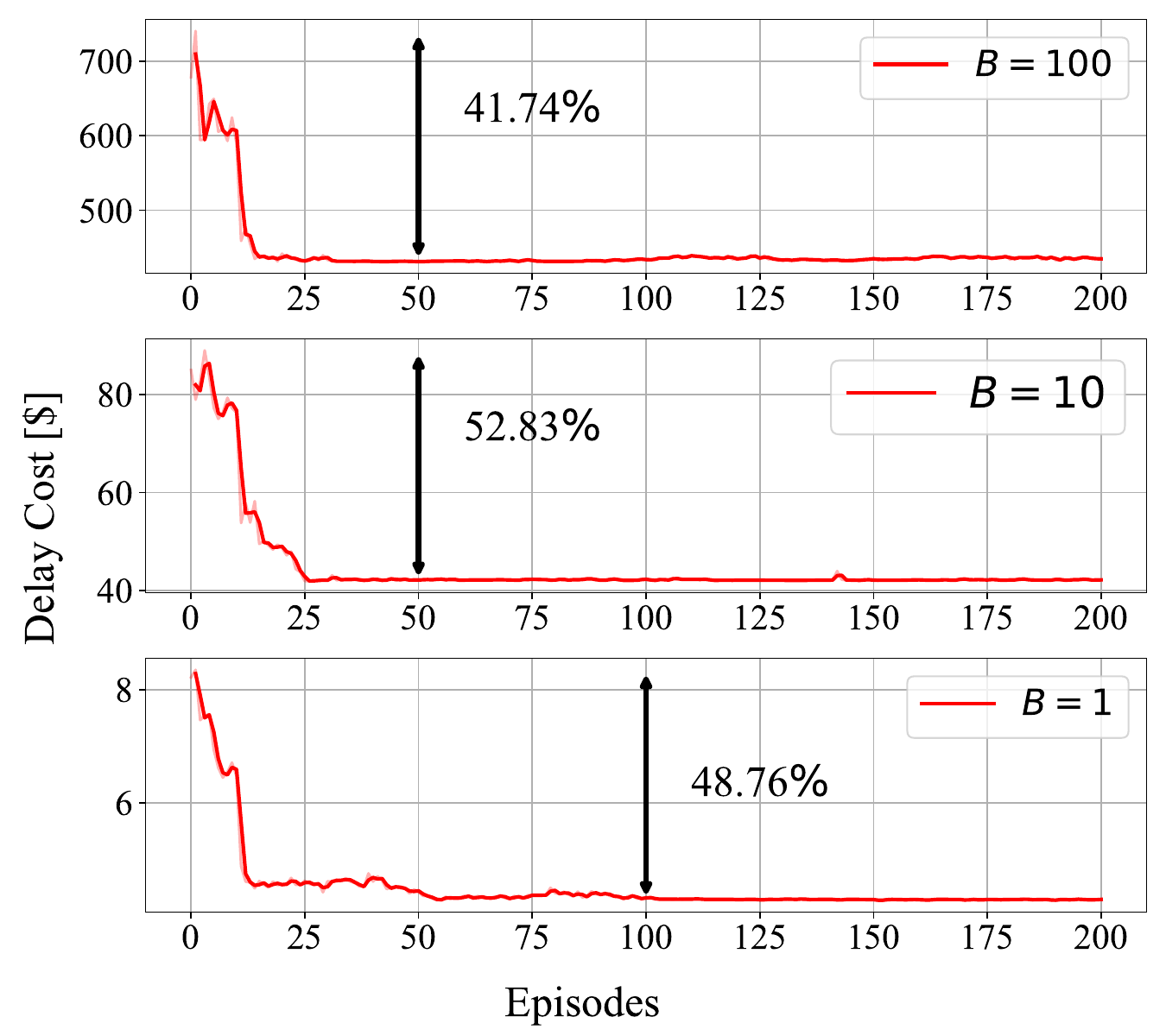}   
	\caption{\small The incurred delay cost over various delay coefficient fees. }	
	\label{fig:delay_cost}
\end{figure}

\subsubsection{Delay minimization} Fig. \ref{fig:delay_cost} shows how our approach successfully reduces the incurred delay cost of delay-elastic MEC services. Overall, the incurred delay cost can be reduced after the agent converges by around 41\%, 52\% and 48\% when $B=100$, $B=10$, and $B=1$, respectively. Our findings also indicate that the rate at which the cost decreases depends on the coefficient for the delay cost. \textit{When the coefficient is high ($B=100$), the delay cost is potentially larger and can produce a much negative reward. As a result, our approach aims to minimize this impact quickly.} And this is validated from Fig. \ref{fig:delay_cost}, where the cost-savings start to converge after around 15 episodes. Conversely, when the coefficient is low ($B=1$), the delay potentially produces a lower cost. As a result, our approach prioritizes delay cost reduction to a lesser extent in this case, i.e., the cost starts to converge after 50 episodes.

\begin{figure}[t!]
	\centering
	\includegraphics[width=0.495\textwidth]{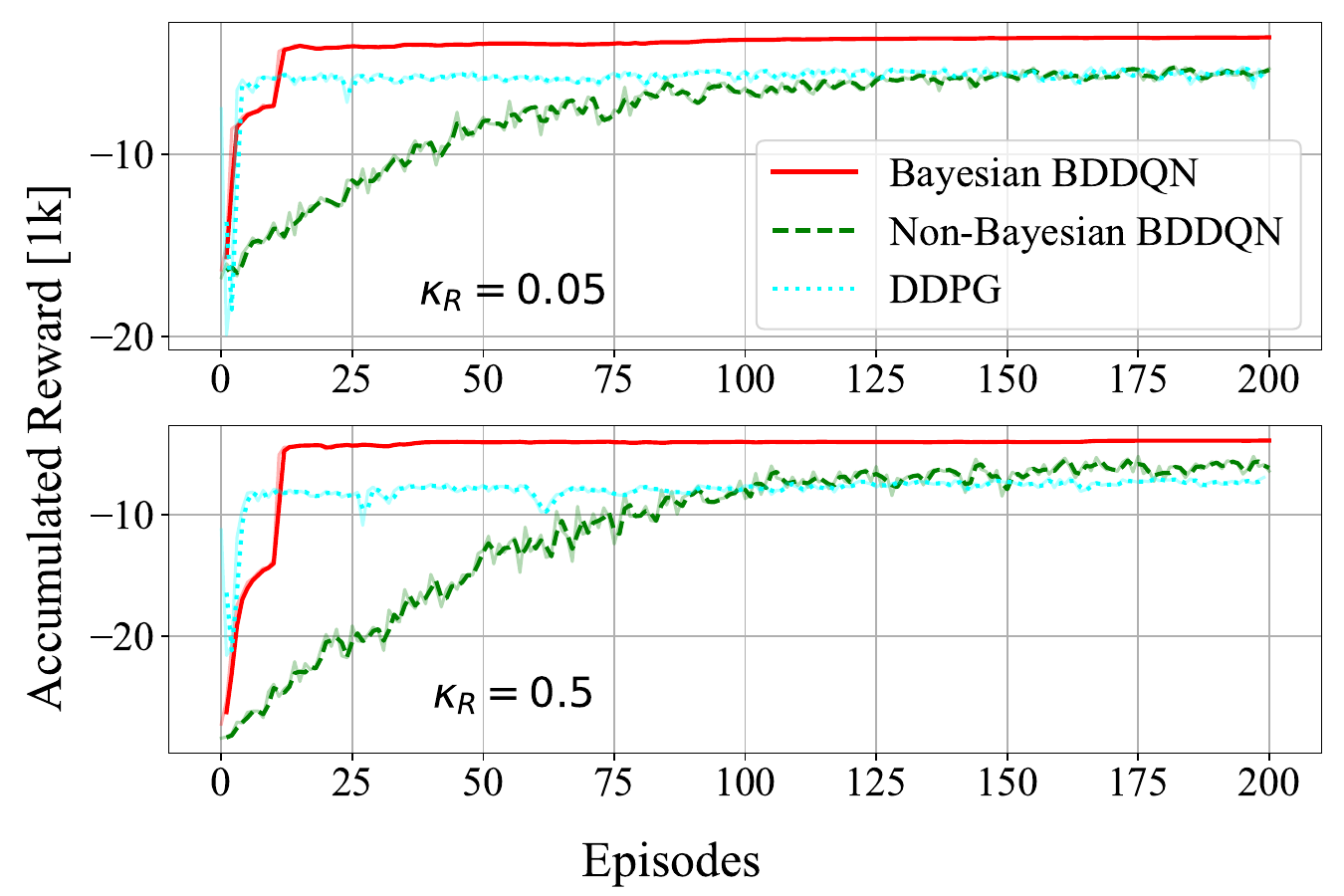}   
	\caption{\small The learning performance of Bayesian BDDQN compared to other benchmarks over two different reconfiguration fees. }	
	\label{fig:reward_r}
\end{figure}

\subsubsection{Impact of reconfiguration fee}

Fig. \ref{fig:reward_r} depicts the learning performance of Bayesian BDDQN over different reconfiguration fees when the pretraining model is unavailable. Regarding reward performance, the results show that Bayesian BDDQN outperforms its non-Bayesian version and DDPG. The performance gap between Bayesian BDDQN and its non-Bayesian version remains at around 25-31\% for both reconfiguration fees. However, compared to DDPG, Bayesian BDDQN performance gain increases with the reconfiguration fees, where it gains 31.12\% for $\kappa_\text{R} = 0.05$ and 44.04\% for $\kappa_\text{R} = 0.5$. \textit{This evaluation highlights that the reconfiguration fee does not significantly affect Bayesian and non-Bayesian BDDQN performance.}

\begin{figure}[t!]
	\centering
	\includegraphics[width=0.48\textwidth]{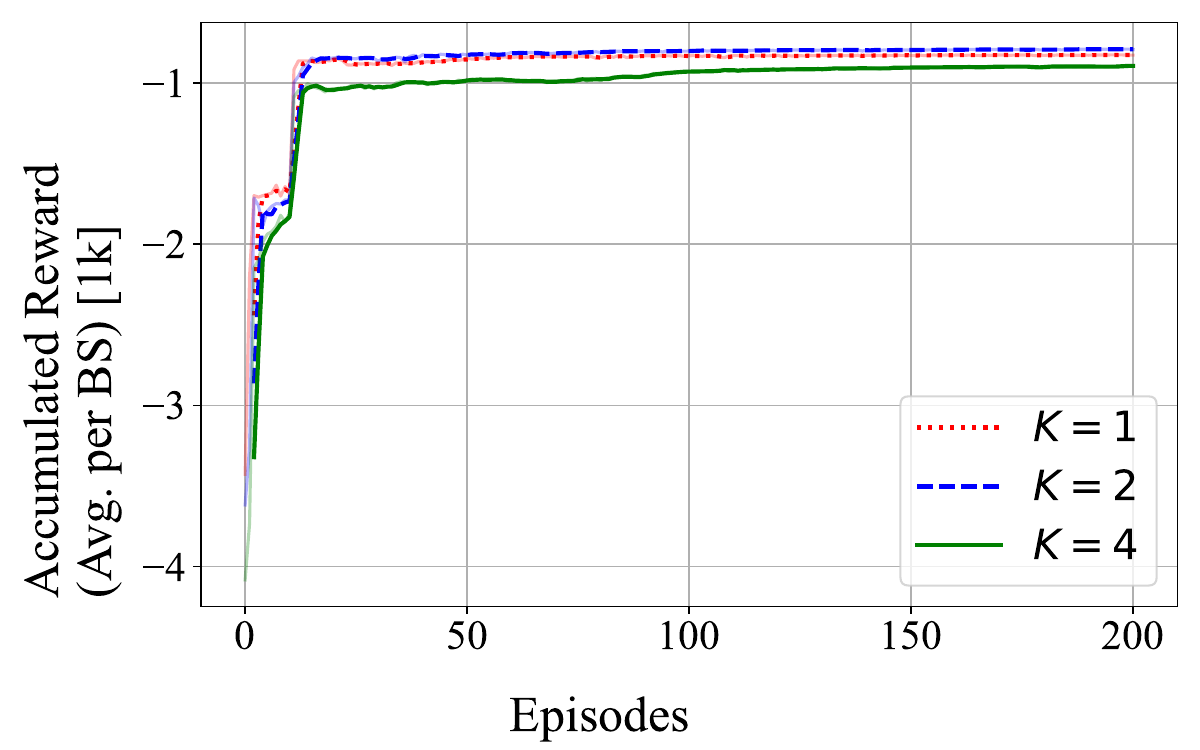}   
	\caption{\small The learning performance of Bayesian BDDQN under different number of BSs and action spaces. }	
	\label{fig:nbs_train}
\end{figure}

\subsubsection{Impact of the number of BSs} We evaluate Bayesian BDDQN over a different number of BSs. The higher number of the BSs indicates the larger action and state spaces in our problem. As illustrated in Fig. \ref{fig:nbs_train}, \textit{Bayesian BDDQN has a similar reward and convergence performance although the size of action space is different.} Some slight differences are found, which indicate that the BS can have different costs than others. When $K=4$, Bayesian BDDQN converges just after 12 episodes and obtains the episodic reward (average per BS) around $-800$. The same trend also appears when $K=1$ and $K=2$.

\begin{figure}[t!]
	\centering
	\includegraphics[width=0.48\textwidth]{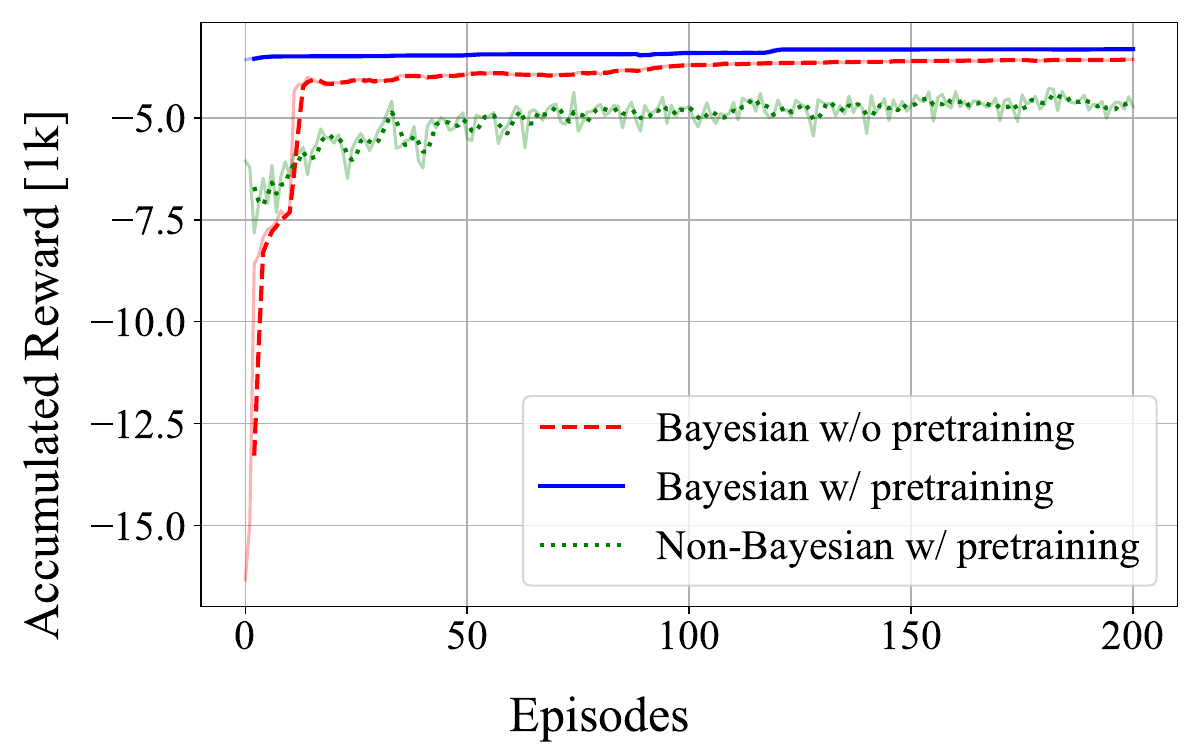}   
	\caption{\small The learning performance of Bayesian BDDQN when a pretraining model is leveraged.  }	
	\label{fig:reward_pretrain}
\end{figure}

\subsubsection{Pretraining models} We evaluate the reward and convergence performance of Bayesian BDDQN when the pretraining model ("Bayesian w/ pretraining") is available. The pretraining model has been trained for 200 episodes in OM1. We utilize it as the weight initialization of Bayesian BDDQN learning, which possibly can further expedite its convergence. The environment for this evaluation is still on OM1, but the input traffic demands are on a different day.
 We use a Bayesian version without pretraining (Bayesian w/o pretraining) and a non-Bayesian version with a pretraining model for benchmarks ("Non-Bayesian w/ pretraining"). For  "Non-Bayesian w/ pretraining", we modify the maximum epsilon parameter from $\epsilon_\text{max} = 1$ into $\epsilon_\text{max} = 0.1$ to encourage less exploration. 

Fig. \ref{fig:reward_pretrain} shows that "Bayesian w/ pretraining" offers the fastest convergence rate and obtains the highest return of the episodic reward compared to the benchmarks. In particular, "Bayesian w/ pretraining" produces a high episodic reward directly at its first learning episode. Then, when the learning goes, it slightly improves the returned reward and eventually outperforms "Bayesian w/o pretraining" and "Non-Bayesian w/ pretraining" by up to 7.02\% and 22.4\%, respectively. Moreover, even leveraging a pretraining model, "Non-Bayesian w/ pretraining" still underperforms the Bayesian approaches. \textit{This evaluation validates that a Bayesian approach can offer performance gain of the non-Bayesian approach even when the pretraining model is unavailable. And this gain can be further increased when a pretraining model is leveraged.}

\begin{figure}[t!]
	\centering
	\includegraphics[width=0.48\textwidth]{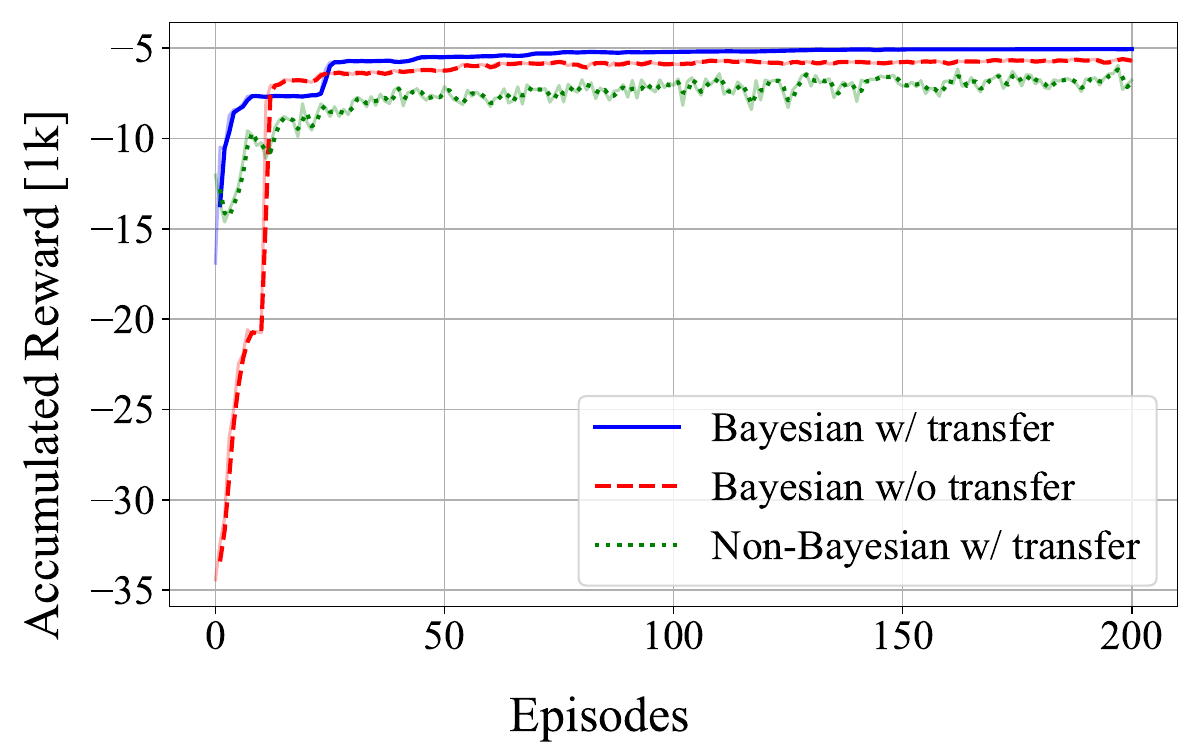}   
	\caption{\small On the performance comparison of Bayesian BDDQN in OM2 when a transfer learning paradigm is adopted. }	
	\label{fig:transfer_training}
\end{figure}

\subsubsection{Transfer Learning} We study the performance of Bayesian BDDQN when the pretraining is available and utilize it for transfer learning ("Bayesian w/ transfer") to a different environment and context. We leverage a pretraining model trained over 200 episodes in OM and use it for the weight initialization of Bayesian BDDQN on OM2. For benchmarks, we compare this approach with Bayesian BDDQN without pretraining ("Bayesian w/o transfer") and its non-Bayesian version with pretraining ("Non-Bayesian w/ transfer"). For  "Non-Bayesian w/ pretraining", the maximum epsilon parameter is modified from $\epsilon_\text{max} = 1$ into $\epsilon_\text{max} = 0.1$. 

Fig. \ref{fig:transfer_training} illustrates that, albeit the pretraining model is leveraged from different O-RAN/MEC systems (environments) and demands (context), "Bayesian w/ transfer" can deliver the highest reward than other benchmarks. Moreover, "Bayesian w/ transfer" can converge as soon as the learning goes (e.g., less than five episodes), and then its reward surpasses "Bayesian w/o transfer," which learns directly from OM2. Fig. \ref{fig:transfer_training} also shows that regardless of with/without transfer learning, Bayesian approaches have a better reward performance than non-Bayesian, even with transfer learning. \textit{These findings emphasize the generalization of our proposed approach over heterogeneous O-RAN/MEC systems, where we possibly reuse the existing pretraining models across different O-RAN/MEC systems and contexts. At the same time, it can offer a fast learning convergence and high reward performance. }

\subsubsection{Penalty}

Fig. \ref{fig:penalty} shows the penalized cost as a result of the enforced actions that violate the constraint requirements. For both Bayesian and non-Bayesian versions, it shows that at the beginning of learning episodes, the cost due to constraint violation is considerably high when the pretraining model is not utilized. One main reason is that it needs some explorations, which renders numerous constraint violations and a severe penalty cost. In contrast, when the pretraining model is adopted, the approaches do not require considerable exploration as they have already acquired some experiences from prior training, which can be valuable in the current learning process. \textit{However, albeit a pretraining model is not available, Bayesian BDDQN can offer fast penalty cost reduction, and eventually, the cost can reach near zero or zero. Moreover, when a pretraining model is leveraged, Bayesian BDDQN can deliver a further advantage, where the penalty cost can reach near zero or zero starting at the beginning of episodes.}

\begin{figure}[t!]
	\centering
	\includegraphics[width=0.48\textwidth]{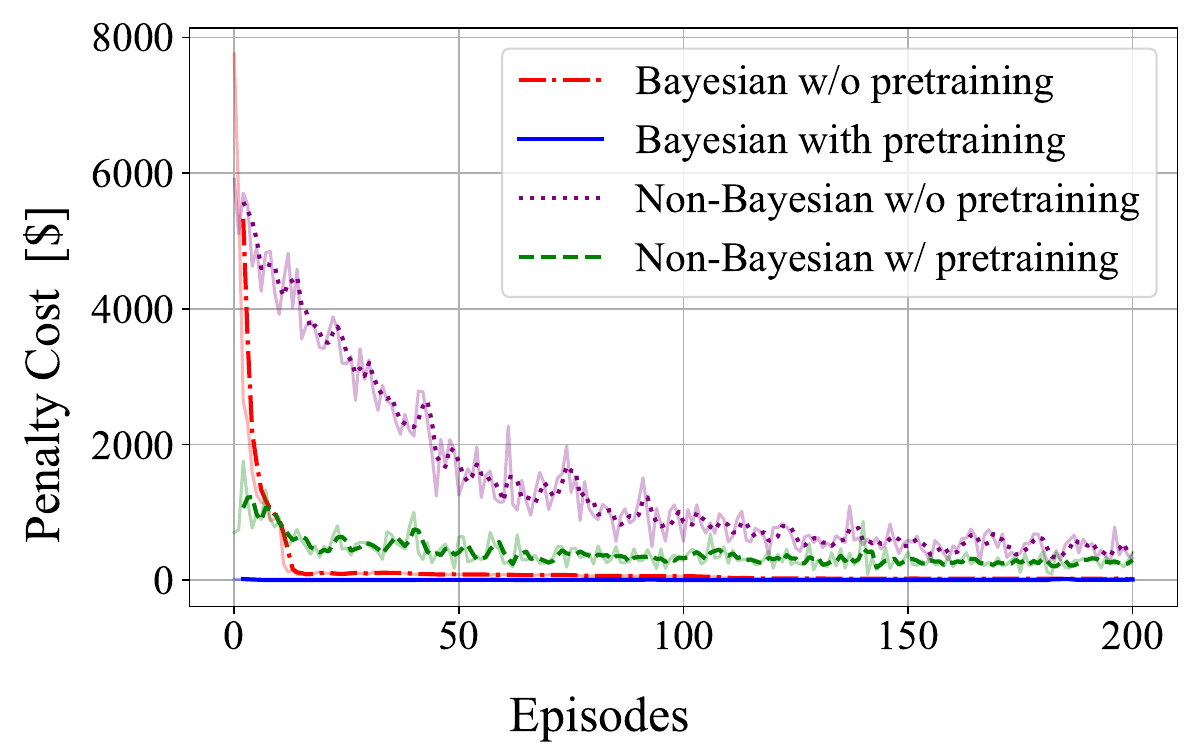}   
	\caption{\small The incurred penalty cost due to constraint violation. }	
	\label{fig:penalty}
\end{figure}


\section{Conclusion} \label{sec:conclusion}
In this paper, we have proposed a fresh O-RAN/MEC orchestration by dynamically controlling the split selection for each BS, the allocated resources for each DU/CU and MEC service, placement for each DU/CU and MEC service over geo-distributed infrastructures, and routing for each legacy/MEC data flow. And the objective is to minimize the long-term overall network operation cost and maximize the MEC performance criterion while adapting possibly time-varying O-RAN/MEC demands and resource availability. We have proposed Bayesian BDDQN for the solution framework based on model free RL paradigm. We developed this framework using a combination of DDQN and action branching, BDDQN, to tackle the large state space and multi-dimensional action space. Further, we tailor a Bayesian framework-based Thompson sampling into BDDQN to encourage data-efficient exploration and improve learning performance. The numerical results have shown that Bayesian BDDQN is provably data-efficient, where it converges faster and improves the learning performance by up to 32\% than its non-Bayesian counterpart, and at the same time, it brings the cost-saving benefits by 41\% compared to DDPG.



{\scriptsize
	\bibliographystyle{IEEEtran}
	\bibliography{IEEEabrv,ref_journal}
}

\end{document}